\documentclass[12pt]{article}
\pdfoutput=1

\usepackage{textcomp}
\usepackage{color}

\usepackage{putex}
\usepackage{autobreak}
\usepackage{graphicx}
\graphicspath{{plots/}}
\usepackage{tabularx}
\usepackage[font=small]{caption}
\usepackage{amsmath}
\usepackage{array}
\usepackage{subcaption}
\usepackage{epstopdf}
\usepackage{enumerate}
\usepackage{cite}
\usepackage{youngtab}
\usepackage{tensor}
\usepackage{slashed}
\usepackage[aligntableaux=center]{ytableau}
\usepackage[utf8]{inputenc}
\usepackage{rotating}
\usepackage{multirow}
\usepackage[
      colorlinks=true,
      linkcolor=blue,
      urlcolor=blue,
      filecolor=black,
      citecolor=red,
      ]{hyperref}
\usepackage{bm}
\usepackage{tikz}

\newcommand{\grp}[1]{\mathrm{#1}}

\newcommand{\grSU}{\grp{SU}}

\newcommand{\grUSp}{\grp{USp}}

\newcommand{\grSL}{SL}

\newcommand{\abs}[1]{\left\lvert #1 \right\rvert}

\newcommand {\be} {\begin {equation}}
\newcommand {\ee} {\end {equation}}

\newcommand {\bes} {\begin {equation*}}
\newcommand {\ees} {\end {equation*}}

\newcommand{\es}[2] {\begin{equation} \label{#1} \begin{split} #2 \end{split} \end{equation}}

\newcommand{\Z}{\mathbb{Z}}

\newcommand{\cF}{{\mathcal F}}

\newcommand{\cO}{{\mathcal O}}

\newcommand{\cT}{{\mathcal T}}

\newcommand{\beq}{\begin{equation}}
\newcommand{\eeq}{\end{equation}}

\def\ie{\begin{equation}\begin{aligned}}
\def\fe{\end{aligned}\end{equation}}

\numberwithin{equation}{section}

\def\<{\langle}
\def\>{\rangle}

\definecolor{mypurple}{rgb}{.5,0,.5}

\newcommand{\gym}{g_\text{YM}}

\graphicspath{{plots/}}

\begin{document}

\preprint{PUPT-2650}

\institution{Imp}{Blackett Laboratory, Imperial College, Prince Consort Road, London, SW7 2AZ, U.K.}
\institution{PU}{Joseph Henry Laboratories, Princeton University, Princeton, NJ 08544, USA}
\institution{PCTS}{Princeton Center for Theoretical Science, Princeton University, Princeton, NJ 08544, USA}
\institution{IAS}{Institute for Advanced Study, Princeton, NJ 08540, USA}

\title{
Level Repulsion in $\mathcal{N} = 4$ super-Yang-Mills via Integrability, Holography, and the Bootstrap
}

\authors{Shai M.~Chester,\worksat{\Imp} Ross Dempsey,\worksat{\PU} and Silviu S. Pufu\worksat{\PU,\PCTS,\IAS}  }

\abstract{We combine supersymmetric localization with the numerical conformal bootstrap to bound the scaling dimension and OPE coefficient of the lowest-dimension operator in $\mathcal{N} = 4$ $\grSU(N)$ super-Yang-Mills theory for a wide range of $N$ and Yang-Mills couplings $g_\text{YM}$. We find that our bounds are approximately saturated by weak coupling results at small $g_\text{YM}$. Furthermore, at large $N$ our bounds interpolate between integrability results for the Konishi operator at small $g_\text{YM}$ and strong-coupling results, including the first few stringy corrections, for the lowest-dimension double-trace operator at large $g_\text{YM}$. In particular, our scaling dimension bounds describe the level splitting between the single- and double-trace operators at intermediate coupling.}
\date{December 2023}

\maketitle

\tableofcontents

\section{Introduction and Summary}
\label{intro}

A central outstanding challenge in Quantum Field Theory is to understand how the spectra of gauge theories change from weak to strong coupling. In Quantum Chromodynamics, this is the question of how a theory of quarks and gluons in the weakly-coupled ultraviolet limit becomes a theory of hadrons in the strongly-coupled infrared limit. A rich model for this kind of reorganization of a gauge theory spectrum is $\mathcal{N}=4$ super-Yang-Mills (SYM) theory. In this paper, we will study this theory with gauge group $\grSU(N)$ and complexified gauge coupling $\tau = \frac{\theta}{2\pi} + \frac{4\pi i}{g_\text{YM}^2}$, where $g_\text{YM}$ is the Yang-Mills coupling and $\theta$ is the theta angle. (In our numerical study we set $\theta = 0$ for simplicity.) $\mathcal{N}=4$ SYM theory is a conformal field theory (CFT) \cite{Grisaru:1980nk,Grisaru:1980jc,Caswell:1980ru,Caswell:1980yi,Sohnius:1981sn}, so the quantities of interest include the scaling dimensions and operator product expansion (OPE) coefficients (referred to below as the CFT data) of its various operators.  In general, these quantities depend non-trivially on $\tau$, with the exception of the scaling dimensions of supersymmetry-protected operators that belong to shortened representations of the superconformal algebra \cite{Minwalla:1997ka,Dolan:2002zh,Cordova:2016emh} and also of some of the OPE coefficients of these operators \cite{Lee:1998bxa}.  As an example, the operator with the overall lowest scaling dimension is the superconformal primary of the multiplet in which the stress-energy tensor resides, and its protected scaling dimension, $\Delta=2$, is independent of $\tau$.  This operator transforms as the ${\bf 20}'$ representation of the $\grSU(4)_R$ R-symmetry.  For small $g_\text{YM}$, the lowest {\em unprotected} operator is the Konishi operator \cite{Konishi:1983hf}.  It is an $\grSU(4)_R$ singlet, and its scaling dimension starts at $\Delta=2+O(g_\text{YM}^2)$ at weak coupling, with an expansion in $g_\text{YM}^2$ that is known up to four-loop order \cite{Velizhanin:2009gv,Eden:2012rr,Fleury:2019ydf,Eden:2016aqo,Goncalves:2016vir}. 

$\mathcal{N}=4$ SYM theory also arises in one of the best-understood instances of the AdS/CFT correspondence \cite{Maldacena:1997re,Gubser:1998bc,Witten:1998qj} (see \cite{Aharony:1999ti,DHoker:2002nbb,Klebanov:2000me,Polchinski:2010hw} for reviews), which relates it to type IIB string theory on $\text{AdS}_5\times S^5$ with complexified string coupling $\tau_s = \chi + i/g_s$, and
\es{adscft}{
 \frac{L^4}{\ell_s^4}=g_\text{YM}^2 N\,,\qquad\tau=\tau_s\,,
}
where $\ell_s$ is the string length, and where AdS$_5$ and $S^5$ both have curvature radii equal to $L$.   The relations \eqref{adscft} imply that $\mathcal{N} = 4$ SYM is described by weakly-curved supergravity at large $N$ and large 't Hooft coupling $\lambda \equiv g_\text{YM}^2 N$, with the $1/N$ and $1/\lambda$ corrections arising from loop and stringy corrections to classical type IIB supergravity.  At large $N$ and $\lambda$, the low-lying single-trace operators of $\mathcal{N} = 4$ SYM theory are dual to fluctuations around $\text{AdS}_5\times S^5$ \cite{Kim:1985ez} and they all belong to short superconformal multiplets with protected scaling dimensions.  By contrast, the unprotected single-trace operators are dual to massive string states.  Their masses are of order $1/\ell_s$, and consequently the AdS/CFT dictionary \cite{Maldacena:1997re,Gubser:1998bc,Witten:1998qj} implies that the scaling dimensions of these operators are of order $\lambda^{1/4}$ \cite{Gubser:2002tv}. The dimensions of the multi-trace operators are just the sums of their single-trace constituents due to large $N$ factorization \cite{tHooft:1973alw}.

\begin{figure}
	\centering
	\includegraphics[width=.8\linewidth]{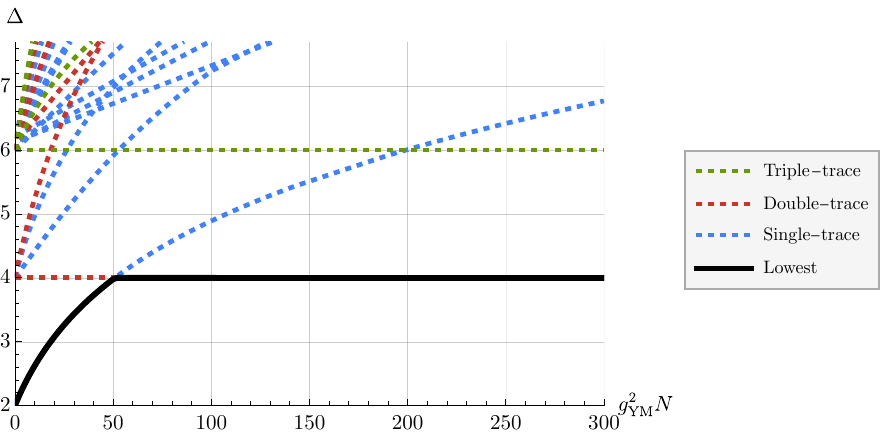}
	\caption{The low-lying spectrum (up to twist six in the free theory) of unprotected scalar superconformal primary operators in $\mathcal{N} = 4$ SYM theory, obtained in the planar limit from the Quantum Spectral Curve as described in Section \ref{4point} \cite{Gromov:2023hzc}. At weak coupling, the lowest unprotected scalar operator is the single-trace Konishi. This operator crosses with a double-trace twist-four operator, which becomes the lowest unprotected scalar at strong coupling.}
	\label{fig:integrability_spectrum}
\end{figure}

An important question in holography is how exactly the spectrum of the weakly-coupled gauge theory, which is rather dense, rearranges itself into that of the weakly-curved supergravity, which by comparison is rather sparse. This question can be answered fully in the planar limit, i.e.~at leading order in large $N$ and at fixed $\lambda$.  In particular, as we increase~$\lambda$, the scaling dimensions of the unprotected single-trace operators  and their multi-trace composites cross those of the protected single-trace operators  and their multi-trace composites and go to infinity as $\lambda^{1/4}$ at large $\lambda$.  This picture can be made quantitatively precise using the integrability of planar $\mathcal{N} = 4$ SYM theory \cite{Minahan:2002ve,Bena:2003wd,Beisert:2003tq,Kazakov:2004qf,Beisert:2005bm,Staudacher:2004tk,Beisert:2005tm,Arutyunov:2004vx,Beisert:2006ib,Beisert:2010jr,Gromov:2023hzc}.\footnote{The single-trace spectrum was also recently used to learn about other aspects of the classical string such as the Hagedorn temperature \cite{Harmark:2017yrv,Urbach:2022xzw,Harmark:2021qma,Bigazzi:2023oqm,Ekhammar:2023glu,Bigazzi:2023hxt} and the AdS Virasoro-Shapiro amplitude \cite{Alday:2023mvu,Alday:2023jdk,Alday:2023flc,Alday:2022xwz,Alday:2022uxp}.}   In Figure~\ref{fig:integrability_spectrum}, based on the integrability results in \cite{Gromov:2023hzc}, we plot the scaling dimensions of the unprotected R-symmetry singlet scalar operators that have twist at most six in the free theory, showing the first few level crossings. As we move away from the planar limit, these level crossings are replaced by level repulsions, and the CFT data are also adjusted due to $1/N$ corrections.  However, it has been difficult to go beyond the planar limit using integrability,\footnote{See \cite{Bargheer:2017nne} for some first steps in this direction.} or even to compute OPE coefficients in the planar limit at finite $\lambda$.\footnote{Planar OPE coefficients were recently computed using integrability for large charge operators \cite{Basso:2015zoa}. The spectrum of scaling dimensions has also been combined with crossing constraints to try to fix OPE coefficients of general operators, which has been very successful for defect CFT
\cite{Cavaglia:2022qpg,Cavaglia:2021bnz,Cavaglia:2022yvv}, but more challenging for the full SYM theory \cite{Caron-Huot:2022sdy}.} Thus, no precise picture that includes level repulsion and $1/N$ corrections is currently available.  (See, however, \cite{Korchemsky:2015cyx} for an approximate treatment.)
 
In this paper, we use a numerical bootstrap\footnote{See \cite{Chester:2019wfx,Poland:2018epd,Poland:2022qrs,Rychkov:2023wsd} for some review of the conformal bootstrap, following the original work \cite{Rattazzi:2008pe}.} approach to investigate both of these effects for {\em the lowest unprotected operator}, which, as seen in Figure~\ref{fig:integrability_spectrum}, interpolates between the Konishi operator at weak coupling and a double-trace composite of the protected $\Delta=2$ operator mentioned above at strong coupling. Specifically, for a wide range of $N$ and $g_\text{YM}$, we bound the scaling dimension of this operator and the OPE coefficient with which it appears in the OPE of two stress tensor multiplets.  As we increase $N$, we find that our scaling dimension bounds approach the planar curve in Figure~\ref{fig:integrability_spectrum}, albeit slowly in the strong-coupling regime. Our bounds are better approximated when we include non-planar corrections at strong coupling obtained from a holographic large-$N$ expansion \cite{Binder:2019jwn,Chester:2019jas,Chester:2020vyz,Chester:2020dja,Chester:2019pvm} derived from analytic bootstrap \cite{Heemskerk:2009pn,Rastelli:2017udc} and supersymmetric localization results \cite{Pestun:2007rz}.  After accounting for these corrections, we find that our bounds smoothly interpolate between the weak-coupling expansion for the Konishi scaling dimension and the large-$N$ expansion for the double-trace scaling dimension, thus providing a non-perturbative completion of these two results that captures the level repulsion between two operators. Our OPE coefficient bounds are also suggestive of this level repulsion.  Based on the fact that our CFT data bounds match the weak and strong coupling analytical approximations,\footnote{This is distinct from the original $\mathcal{N} = 4$ bootstrap study \cite{Beem:2013qxa}, for which one can check that, at least at order $1/c^2$, where $c$ is the conformal anomaly coefficient, the scaling dimension bounds are nearly saturated by a putative pure AdS$_5$ supergravity theory and not type IIB string theory on AdS$_5 \times S^5$ \cite{Alday:2022ldo}.} we conjecture that they come close to being saturated by ${\cal N}=4$ SYM for all values of $N$ and $g_\text{YM}$.\footnote{In many other bootstrap studies, the numerical bootstrap bounds were also found to be nearly saturated by physical theories such as the critical $O(N)$ vector models \cite{ElShowk:2012ht,Kos:2013tga,Kos:2015mba,Kos:2016ysd,Chester:2020iyt,Chester:2019ifh}, QED$_3$ \cite{Chester:2016wrc,Albayrak:2021xtd,Chester:2023njo}, and the 3-state Potts model \cite{Chester:2022hzt}.}

As in the original numerical bootstrap study of ${\cal N} = 4$ SYM in \cite{Beem:2013qxa,Beem:2016wfs}, in order to derive our scaling dimension and OPE coefficient bounds, we impose constraints coming from supersymmetry and crossing symmetry on the four-point function of four stress tensor multiplet operators.\footnote{Superconformal symmetry implies that the four-point functions of any four operators from these multiplets are algebraically related to the four-point function of the superconformal primary \cite{Belitsky:2014zha}.}  The value of $N$ can be inputted through the conformal anomaly or using the values of protected OPE coefficients derived from the 2d chiral algebra \cite{Beem:2013sza}. In addition, as in our previous work \cite{Chester:2021aun}, we also impose constraints that require certain integrals of the stress tensor four-point function to take values that can be computed using supersymmetric localization.  The latter constraints are the only ones that are sensitive to the gauge coupling $\tau$.\footnote{As shown in \cite{Chester:2021aun}, the $\tau$-independent bounds of \cite{Beem:2013qxa,Beem:2016wfs} are strictly less stringent than the $\tau$-dependent bounds obtained using the additional supersymmetric localization constraint.}  In particular, certain integrals of the stress tensor four-point function were related in \cite{Binder:2019jwn,Chester:2020dja} to mass derivatives $\partial_m^4F\vert_{m=0}$ and $\partial_\tau\partial_{\bar\tau}\partial_m^2F\vert_{m=0}$ of the mass-deformed $S^4$ free energy $F(m, \tau, \bar \tau)$, which can be computed using supersymmetric localization as an exact function of $\tau$ in terms of an $(N-1)$-dimensional integral \cite{Pestun:2007rz}.  The difference between the present work and \cite{Chester:2021aun} is that in \cite{Chester:2021aun} the numerical bootstrap approach was limited to $N=2$ and $3$ due to the fact that the $(N-1)$-dimensional integral was difficult to evaluate for large $N$, while here we extend the analysis to $N>3$.\footnote{We also introduce a slight variation of the numerical method in \cite{Chester:2021aun} that is more numerically stable at higher bootstrap precision.}   This extension is possible due to the observation \cite{Alday:2023pet} that both $\partial_m^4F\vert_{m=0}$ and $\partial_\tau\partial_{\bar\tau}\partial_m^2F\vert_{m=0}$ can be accurately computed for all $N$ and $\tau$ in terms of just two integrals, by combining the large-$N$, finite-$\tau$ expansion in \cite{Chester:2019jas,Chester:2020vyz} with exact results for the non-instanton sector \cite{Chester:2019pvm,Chester:2020dja}.

The rest of this paper is organized as follows. In Section~\ref{4point} we review the conformal block decomposition of the stress tensor four-point correlator, the method by which localization can be used to constrain the correlator, and the weak-coupling, analytic bootstrap, and integrability results for local CFT data. In Section~\ref{numBoot}, we introduce a slight variation of the integration domain for the integrated constraints that is more numerically stable when combined with the numerical bootstrap, and use this algorithm to compute bounds on CFT data for a wide range of $N$ and $\tau$, which we compare to the various perturbative results. Lastly, in Section~\ref{disc} we end with a discussion of our results and of future directions. Various technical details are discussed in the Appendix.

\section{Stress tensor correlator}
\label{4point}

Our numerical bootstrap analysis relies on consistency conditions imposed on the four-point correlator
 \es{Correlator}{
  \langle S_{I_1 J_1}(\vec{x}_1)S_{I_2 J_2}(\vec{x}_2)S_{I_3 J_3}(\vec{x}_3)S_{I_4 J_4}(\vec{x}_4)\rangle
 }  
 of the superconformal primary $S_{IJ}(\vec{x})$ of the stress tensor multiplet of the ${\cal N}=4$ SYM theory.  Here, $\vec{x}$ is a spacetime coordinate, the indices $I, J$ run from $1$ to $6$, and $S$ is a symmetric traceless tensor, as is appropriate for the ${\bf 20}'$ representation of $\grSU(4)_R$.   As explained in detail in \cite{Beem:2016wfs,Beem:2013qxa,Chester:2021aun}, superconformal symmetry implies that \eqref{Correlator} can be written in terms of a single function $\cT(U,V)$ of the conformal cross-ratios 
 \es{CrossRatios}{
   U \equiv \frac{{x}_{12}^2 {x}_{34}^2}{{x}_{13}^2 {x}_{24}^2} \,, \qquad 
     V \equiv \frac{{x}_{14}^2 {x}_{23}^2}{{x}_{13}^2 {x}_{24}^2} \,,
  }
with $\vec{x}_{ij} \equiv \vec{x}_i - \vec{x}_j$. In this section we start with a brief review of just the properties of the function $\cT(U,V)$ that are needed for the numerical bootstrap analysis.  For additional background, we refer the reader to \cite{Chester:2021aun}  (and references therein), whose notation we follow. Afterwards, we discuss perturbative results at small $\lambda\equiv g_\text{YM}^2 N$ and finite $N$, at large $N$ and finite $\lambda$ from integrability, and at large $N$ and finite $\tau$ from the analytic bootstap combined with localization.

\subsection{Properties of $\cT(U, V)$}
\label{setup}

The first property of $\cT$ we will need is its transformation under the crossing symmetry of the four-point function \eqref{Correlator}, which requires it to obey the two relations (see for instance (3.31) of \cite{Chester:2020dja})\footnote{One can obtain three more similar relations by combining the two relations in \eqref{Crossing}.  All five relations are given explicitly in (3.31) of \cite{Chester:2020dja}.}
 \es{Crossing}{
  \cT(V, U) = \frac{V^2}{U^2} \cT(U, V) \,, \qquad \cT\left(\frac UV , \frac 1V \right) = V^2 \cT(U, V) \,.
 }
More abstractly, crossing symmetry corresponds to permuting the insertion points $\vec{x}_i$ of the four operators.  Such permutations are isomorphic to the symmetric group $S_4$.  However, since the even permutations that replace $(\vec{x}_1, \vec{x}_2, \vec{x}_3, \vec{x}_4)$ by either $(\vec{x}_2, \vec{x}_1, \vec{x}_4, \vec{x}_3)$, $(\vec{x}_3, \vec{x}_4, \vec{x}_1, \vec{x}_2)$, or $(\vec{x}_4, \vec{x}_3, \vec{x}_2, \vec{x}_1)$ are the non-trivial elements of a $\Z_2 \times \Z_2$ normal subgroup of $S_4$ that leaves the cross-ratios $(U, V)$ invariant, only the subgroup $S_4/(\Z_2 \times \Z_2) = S_3 \subset S_4$ acts non-trivially on $(U, V)$.  This $S_3$ subgroup can be represented by permutations of only $(\vec{x}_1, \vec{x}_2, \vec{x}_3)$ with fixed $\vec{x}_4$.  The crossing symmetry \eqref{Crossing} of the function ${\cal T}$ can be restated more succinctly by saying that $\tilde \cT(U, V) \equiv \cT(U,V) U^{-2/3} V^{4/3}$ is $S_3$-invariant in the sense that $\tilde \cT(U, V) = \tilde \cT(U', V')$ for any $(U, V)$ and $(U', V')$ related by an $S_3$ permutation.  Indeed, the two relations exhibited in \eqref{Crossing} are obtained from the invariance of $\tilde \cT$ under the transpositions $(\vec{x}_1, \vec{x}_2, \vec{x}_3) \to (\vec{x}_3, \vec{x}_2, \vec{x}_1)$ and $(\vec{x}_1, \vec{x}_2, \vec{x}_3) \to (\vec{x}_2, \vec{x}_1, \vec{x}_3)$.  Since these two transpositions generate the entire $S_3$ crossing symmetry group, the two independent crossing relations obeyed by $\cT$ can be taken to be those in \eqref{Crossing}.

The second property we will need corresponds to the decomposition of \eqref{Correlator} into superconformal blocks.  With respect to the $s$-channel OPE, $\cT$ can be split as
 \es{TExp}{
  \cT(U,V) = \cT_\text{long}(U,V) + \cT_\text{short}(U,V) \,,
 }
where $ \cT_\text{short}$ is defined in \eqref{Fshort} and, in the $s$-channel OPE, receives contributions only from short representations of the superconformal algebra, while $ \cT_\text{long}$ receives contributions only from long representations.   $\cT_\text{long}$ can be expanded as
 \es{Gexp}{
 \cT_\text{long}(U,V)=\frac{1}{U^2} \sum_{{\cal O}_{\Delta, \ell}} \lambda^2_{\Delta,\ell}G_{\Delta+4,\ell}(U,V)\,,
 }
 where each term in the sum corresponds to the exchange of a superconformal primary ${\cal O}_{\Delta, \ell}$ of spin $\ell=0, 2, 4, \ldots$ and scaling dimension $\Delta \geq \ell+2$ together with all of its superconformal descendants, the $\lambda^2_{\Delta,\ell}$ are squares of OPE coefficients, and $G_{\Delta,\ell}(U,V)$ are 4d conformal blocks with scaling dimension $\Delta$ and spin $\ell$, namely
 \es{4dblock}{
 G_{\Delta,\ell}(U,V) &=\frac{z\bar z}{z-\bar z}(k_{\Delta+\ell}(z)k_{\Delta-\ell-2}(\bar z)-k_{\Delta+\ell}(\bar z)k_{\Delta-\ell-2}( z))\,,\\
k_h(z)&\equiv z^{\frac h2}{}_2F_1(h/2,h/2,h,z)\,,
 }
where $U=z \bar z$ and $V=(1-z)(1-\bar z)$. 

\begin{figure}
	\centering
	\begin{subfigure}[t]{0.48\textwidth}
		\centering
		\includegraphics[width=\linewidth]{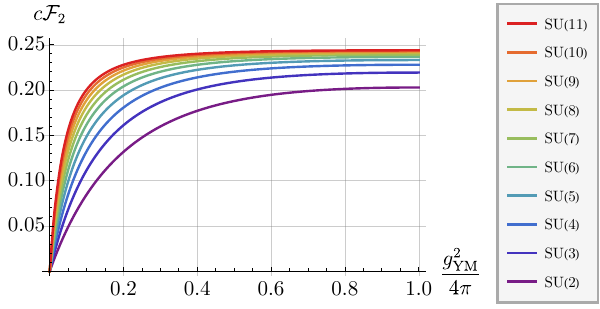}
		\caption{}
		\label{fig:localization_m2}
	\end{subfigure}%
	\hspace{.02\textwidth}
	\begin{subfigure}[t]{0.48\textwidth}
		\centering
		\includegraphics[width=\linewidth]{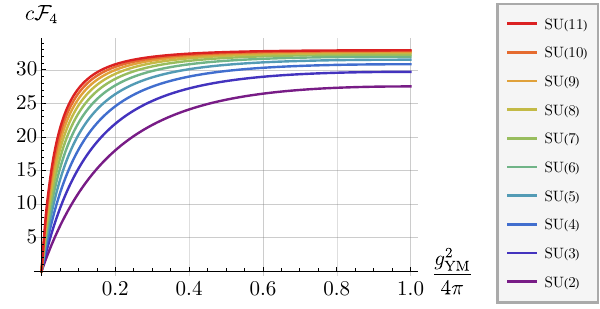}
		\caption{}
		\label{fig:localization_m4}
	\end{subfigure}%
	\caption{The integrals defined in \eqref{ints}, rescaled by $c = \frac{N^2 - 1}{4}$ and evaluated as a function of $g_\text{YM}^2 N$ for $N = 2$ through $11$.}
	\label{fig:localization_fd}
\end{figure}

The last property of $\cT$ we will use is that certain integrals of it can be calculated using supersymmetric localization. In particular, if  $F(m,\tau,\bar\tau)$ is the mass-deformed sphere free energy,  then \cite{Binder:2019jwn,Chester:2020dja}
 \es{constraint1}{
\mathcal{F}_2(\tau,\bar\tau)\equiv\frac{1}{8c}\frac{\partial_m^2\partial_\tau\partial_{\bar\tau} F}{\partial_\tau\partial_{\bar\tau} F}\Big\vert_{m=0} &=I_2\left[\cT\right] \,,\\
\mathcal{F}_4(\tau,\bar\tau)\equiv 48 \zeta(3) c^{-1}+ c^{-2}{\partial^4_m  F}\big\vert_{m=0}  &=I_4\left[ \cT \right]\,,
 }
where $c \equiv \frac{N^2 - 1}{4}$ is the conformal anomaly coefficient and the integrals are defined as
  \es{ints}{
   I_{2}[f]&\equiv -\frac{2}{\pi}\int dR\, d\theta\, \frac{R^3\sin^2\theta f(U,V)}{U^2}\bigg|_{\substack{U = 1 + R^2 - 2 R \cos \theta \\
    V = R^2}}\,,\\
       I_4[f]&\equiv-\frac{32}{\pi}  \int dR\, d\theta\, R^3 \sin^2 \theta \, (U^{-1}+U^{-2}V+U^{-2})\bar{D}_{1,1,1,1}(U,V) f(U, V) \bigg|_{\substack{U = 1 + R^2 - 2 R \cos \theta \\
    V = R^2}}\,,\\
 }
 and $\bar{D}_{1,1,1,1}(U,V)$ takes the form
 \es{Dbar}{
\bar{D}_{1,1,1,1}(U,V)= \frac{1}{z-\bar z}\left(\log(z \bar z)\log\frac{1-z}{1-\bar z}+2\text{Li}(z)-2\text{Li}(\bar z)\right)\,.
 }
 
The mass derivatives $\cF_2(\tau,\bar\tau)$ and $\cF_4(\tau,\bar\tau)$ are written in terms of the partition function $Z(m,\tau,\bar\tau)\equiv \exp(-F(m,\tau,\bar\tau))$, which was computed using supersymmetric localization in \cite{Pestun:2007rz} for $\grSU(N)$\footnote{Similar expressions are given in \cite{Pestun:2007rz} in terms of other classical groups, but we only consider $\grSU(N)$ here for simplicity.} $\mathcal{N}=4$ SYM in terms of the matrix model integral
  \es{ZFull}{
  Z(m, \tau, \bar \tau) = \int \frac{d^{N-1} a}{N!} \,  \frac{  \prod_{i < j}a_{ij}^2 H^2(a_{ij})}{ H(m)^{N-1} \prod_{i \neq j} H(a_{ij}+ m)}e^{- \frac{8 \pi^2}{g_\text{YM}^2} \sum_i a_i^2} \abs{Z_\text{inst}(m, \tau, a_{ij})}^2 \,,
 }
where $a_{ij}\equiv a_i-a_j$, the integration is over $N$ real variables $a_i$, $i = 1, \ldots, N$ subject to the constraint $\sum_i a_i = 0$, the function $H(m) \equiv e^{-(1 + \gamma) m^2} G(1 + im) G(1 - im)$ is the product of two Barnes G-functions, and $Z_\text{inst}$ is the contribution from instantons localized at the poles of $S^4$, whose explicit form is a complicated infinite sum over instanton sectors as shown in \cite{Nekrasov:2002qd,Nekrasov:2003rj}. For low $N$, the mass derivatives in \eqref{constraint1} can be computed in terms of $N-1$ numerical integrals by truncating the instanton expansion for $Z_\text{inst}$, which converges rapidly in the $SL(2,\mathbb{Z})$ fundamental domain
\es{tauDomain}{
|\tau|\geq1\,,\qquad |\Re(\tau)|\leq\frac12\,.
}
For arbitrary $N$, it was shown in \cite{Alday:2023pet} how $\cF_2(\tau,\bar\tau)$ and $\cF_4(\tau,\bar\tau)$ can be efficiently computed to high precision for any $N$ and $\tau$ by combining the exact expressions for the non-instanton sector as computed using orthogonal polynomials in \cite{Chester:2019pvm,Chester:2020dja} with the instanton sector of the large $N$ finite $\tau$ expansion of \cite{Chester:2019jas,Chester:2020vyz}. For instance, we can approximate $\cF_2(\tau,\bar\tau)$ as
\es{GHdiv}{
\cF_2(\tau,\bar\tau)\approx&-\frac{\tau_2^2}{4c^2}\partial_{\tau_2}^2\int_0^\infty dw\frac{e^{-\frac{w^2}{\pi\tau_2}}}{2\sinh^2 w}\Big[L_{N-1}^{(1)}({\scriptstyle\frac{w^2}{\pi\tau_2}})-\sum_{i,j=1}^N(-1)^{i-j}L_{i-1}^{(j-i)}({\scriptstyle\frac{w^2}{\pi\tau_2}})L_{j-1}^{(i-j)}({\scriptstyle\frac{w^2}{\pi\tau_2}})\Big]\\
+&\frac{1}{4c^2}\Bigg[
 -\frac{3\sqrt{N}}{2^4 } E( {\scriptstyle {3 \over 2}};\tau, \bar \tau)+\frac{45}{2^8 \sqrt{N}}E( {\scriptstyle {5 \over 2}};\tau, \bar \tau)+\frac{1}{{N}^{\frac32}}\big[-\frac{39}{2^{13} }E( {\scriptstyle {3 \over 2}};\tau, \bar \tau)+\frac{4725}{2^{15}}E( {\scriptstyle {7 \over 2}};\tau, \bar \tau)\big]\\
&\qquad+\frac{1}{{N}^{\frac52}}\big[-\frac{1125}{2^{16} }E( {\scriptstyle {5 \over 2}};\tau, \bar \tau)+\frac{99225}{2^{18} }E( {\scriptstyle {9 \over 2}};\tau, \bar \tau)\big]\Bigg]_{k\neq 0}\,,\\
}
where $\tau\equiv \tau_1+i\tau_2$.
In the first line we have the exact expression for the non-instanton terms from \cite{Chester:2019pvm}, which for all $N$ is written as a single integral of a finite sum of generalized Laguerre polynomials $L_i^{(j)}(x)$. In the other lines, we write the instanton part (denoted by $k\neq0$) of the large $N$ finite $\tau$ expansion from \cite{Chester:2019jas}, which is written in terms of non-holomorphic Eisenstein series 
\es{EisensteinExpansion}{
  E(s, \tau, \bar\tau)
   ={}& {2 \zeta(2 s)}{\tau_2^s} + 2 \sqrt{\pi} \tau_2^{1-s} \frac{\Gamma(s - \frac 12)}{\Gamma(s)} \zeta(2s-1) \\
    &+ \frac{2 \pi^s\sqrt{\tau_2}}{\Gamma(s) } \sum_{k\neq 0} |k|^{s-\frac12}
    \sigma_{1-2s}(|k|) \, 
      K_{r - \frac 12} (2 \pi\tau_2 \abs{k}) \, e^{2 \pi i k\tau_1} \, ,
 }
   where the divisor sum $\sigma_p(k)$ is defined as $\sigma_p(k)=\sum_{d>0,{d|k}}  d^p$, $K_{s - \frac 12}$ is the Bessel function of second kind of index $s-1/2$, and the instanton terms consist of the $k\neq0$ sum. The infinite sum over $k$ converges rapidly in the fundamental domain \eqref{tauDomain}, so as shown in \cite{Alday:2023pet} only a few terms are needed to match the exact expression for $\cF_{2}(\tau,\bar\tau)$ as computed by taking numerical integrals for low $N$. We can similarly approximate $\cF_4(\tau,\bar\tau)$, except the non-instanton expression from \cite{Chester:2020dja} is written in terms of two integrals for all $N$, while the large $N$ and finite $\tau$ expression includes both Eisensteins as well as generalized Eisensteins \cite{Green:2014yxa}. The explicit expression is quite complicated, and is given in the \texttt{Mathematica} notebook attached to \cite{Alday:2023pet}. In Figure~\ref{fig:localization_fd} we plot $\cF_2(\tau,\bar\tau)$ and $\cF_4(\tau,\bar\tau)$ as functions of $\tau$ along the imaginary axis for the various $N$ we will consider in the following sections.

Our numerical bootstrap study combines the conformal block expansion \eqref{Gexp} with the crossing equations \eqref{Crossing} and the integrated constraints \eqref{constraint1}.  While the second crossing equation in \eqref{Crossing} is satisfied automatically by the conformal block expansion, the first equation in \eqref{Crossing} becomes
 \es{CrossAgain}{
   \sum_{{\cal O}_{\Delta, \ell}} \lambda^2_{\Delta,\ell} \left( V^4 G_{\Delta+4,\ell}(U,V) - U^4 G_{\Delta+4,\ell}(U,V)\right)  + U^2 V^4 \cT_\text{short}(U, V) - U^4 V^2 \cT_\text{short}(V, U) = 0 \,.
 }
Similarly, the integrated constraints \eqref{constraint1} become
 \es{IntAgain}{
 \sum_{{\cal O}_{\Delta, \ell}} \lambda^2_{\Delta,\ell} I_2\left[\frac{G_{\Delta+4,\ell}(U,V)}{U^2} \right] + I_2\left[\cT_\text{short} \right] - \cF_2(\tau,\bar\tau)&=0 \,, \\
 \sum_{{\cal O}_{\Delta, \ell}} \lambda^2_{\Delta,\ell} I_4\left[\frac{G_{\Delta+4,\ell}(U,V)}{U^2} \right] + I_4\left[\cT_\text{short} \right] - \cF_4(\tau,\bar\tau)&=0 \,.
 } 
The numerical values of $I_2\left[\cT_\text{short} \right]$ and $I_4\left[\cT_\text{short} \right]$ are given in \eqref{exactInts}.

\subsection{Perturbative expansions}
\label{weak}

The unprotected data in $\mathcal{T}(U, V)$ is all encoded in the scaling dimensions and OPE coefficients of the superconformal primaries $\mathcal{O}_{\Delta,\ell}$ exchanged in the expansion \eqref{Gexp}. These quantities can be calculated in a weak coupling expansion about the free theory point. In the free theory, the lowest unprotected scalar is the Konishi operator, which has twist two. Its scaling dimension and OPE coefficient are known to four-loop order in $\lambda \equiv \gym^2 N$ \cite{Velizhanin:2009gv,Eden:2012rr,Fleury:2019ydf,Eden:2016aqo,Goncalves:2016vir}, and non-planar $\frac{1}{N}$ corrections first appear at order $\lambda^4$. The expansions are 
\es{konishiWeak}{
\Delta_{2,0}&=2+\frac{3 \lambda }{4 \pi ^2}-\frac{3 \lambda ^2}{16 \pi ^4}+\frac{21 \lambda ^3}{256 \pi ^6}+\frac{\lambda ^4
   \left(-1440 \left(\frac{12}{N^2}+1\right) \zeta (5)+576 \zeta (3)-2496\right)}{65536 \pi ^8} + O(\lambda^5)\,,\\
   \lambda_{2,0}^2&=\frac1c\Bigg[\frac{1}{3}-\frac{\lambda }{4 \pi ^2}+\frac{\lambda ^2 (3 \zeta (3)+7)}{32 \pi
   ^4}-\frac{\lambda ^3 (8 \zeta (3)+25 \zeta (5)+48)}{256 \pi ^6}\\
   &\quad+\frac{\lambda ^4 \left(  2488 + 328 \zeta(3) + 72 \zeta(3)^2 + 980 \zeta(5) + 1470 \zeta(7) +\frac{45}{N^2} (8 \zeta
   (5)+7 \zeta (7))\right)}{16384 \pi ^8 }+ O(\lambda^5)\Bigg]\,. \\
}

\begin{figure}[]
\begin{center}
	\begin{subfigure}[t]{0.48\textwidth}
		\centering
        \includegraphics[width=\linewidth]{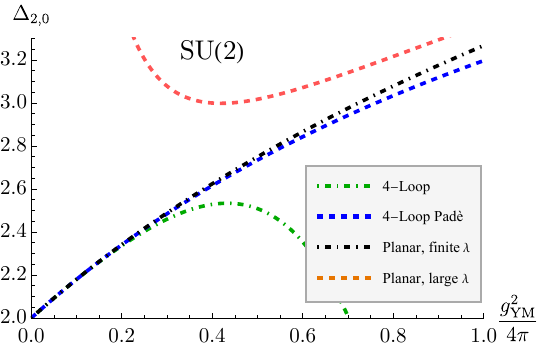}
        \caption{}
    \end{subfigure}%
    \hspace{.02\textwidth}%
    \begin{subfigure}[t]{0.48\textwidth}
		\centering
        \includegraphics[width=\linewidth]{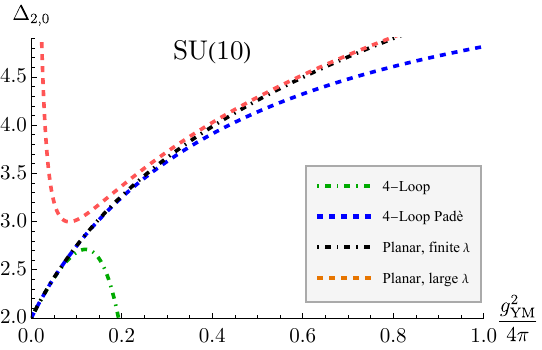}
        \caption{}
        \label{fig:intvsW10}
    \end{subfigure}%
\caption{Large $N$ integrability results for the unique twist two scalar operators $\Delta_{2,0}$  compared for $N=2,10$ to weak coupling estimates at 3 and 4-loop order, as well as with a $(2,2)$ Pad\'e resummation, and also to the large $\lambda$ expansion. We plot in the fundamental domain of the complexified coupling $\tau$ along the imaginary axis from $\tau = \infty$ to $\tau = i$, or equivalenty from $\gym = 0$ to $\gym = \sqrt{4\pi}$.}
\label{fig:intvsW}
\end{center}
\end{figure}  

As discussed in the introduction, the scaling dimensions of single-trace operators can also be computed in the strict planar limit by using the integrability of planar $\mathcal{N} = 4$ SYM theory. The results are encoded in a finite set of equations called the Quantum Spectral Curve \cite{Gromov:2013pga,Gromov:2014caa}. At finite $\lambda$, these equations can be solved numerically \cite{Gromov:2023hzc}, yielding scaling dimensions such as those in Figure \ref{fig:integrability_spectrum}. 

At small and at large $\lambda$, these equations can also be solved analytically. If we solve the equations for the Konishi operator at small $\lambda$, we would recover $\Delta_{2,0}$ as in the weak-coupling expansion, but without the non-planar corrections. Since these corrections only enter at order $\lambda^4$, the integrability results agree very well with the weak-coupling expansion at small $\lambda$. At large $\lambda$, one can also solve the equations of the quantum spectral curve analytically \cite{Gromov:2011bz,Gromov:2014bva,Basso:2011rs,Gromov:2011de}, yielding
\es{largeLam}{
\Delta_{2,0} = 2\lambda^{1/4}-2+\frac{2}{\lambda^{1/4}}+\frac{1/2-3\zeta(3)}{\lambda^{3/4}}+\frac{1/2+6\zeta(3)+15\zeta(5)/2}{\lambda^{5/4}}+O(\lambda^{-3/2})\,.
}
In particular, the leading term indicates that this operator will acquire a very large anomalous dimension at large $\lambda$, and thus cross with the twist-four double-trace operator, as shown in Figure \ref{fig:integrability_spectrum}.

In Figure~\ref{fig:intvsW}, we compare these different estimates for the scaling dimension of the Konishi operator. We look at $N = 2$ and $N = 10$ from $\gym = 0$ to the self-dual point $\gym = \sqrt{4\pi}$, with $\theta = 0$. As we expect, at weak coupling the four-loop expansion agrees well with the numerical integrability result at finite $\lambda$, since the two estimates agree up to $\mathcal{O}(\lambda^3)$. We also plot a (2,2) Pad\'e approximant to \eqref{konishiWeak}, which follows the finite $\lambda$ integrability results for a much larger range of the coupling. At large $\lambda$, we can also compare the expansion \eqref{largeLam} with the finite $\lambda$ integrability results. In Figure \ref{fig:intvsW10}, we see that the match becomes quite good at sufficiently large $\gym^2 N$. However, the two curves only begin to match well around when the scaling dimension exceeds 4, at which point the Konishi is no longer the lowest unprotected scalar operator.

In this regime, the lowest unprotected scalar is instead the twist-four double-trace operator. In the planar limit, the scaling dimension of this operator is exactly 4 at any value of the coupling. Non-planar corrections have been calculated in an expansion at large $N$ with finite $\tau$, by combining the localization constraints \eqref{constraint1} with the analytic bootstrap. For the CFT data of the twist-four double-trace operator, this yields \cite{Chester:2020vyz,Chester:2019jas,Chester:2019pvm,Aprile:2017bgs,Alday:2017xua}
\es{largeN}{
\Delta_{4,0}&=4-\frac{4}{c}+\frac{135}{7\sqrt{2}\pi^{3/2}c^{7/4}}E({\scriptstyle{\frac32}},\tau)+\frac{1199}{42c^2}-\frac{3825}{32\sqrt{2}\pi^{5/2}c^{9/4}}E({\scriptstyle{\frac52}},\tau)+O(c^{-5/2})\,,\\
 \lambda_{4,0}^2&=\frac{1}{10}+\frac{19}{300c}-\frac{4059}{1960\sqrt{2}\pi^{3/2}c^{7/4}}E({\scriptstyle{\frac32}},\tau)+\frac{1}{c^2}\left[a-\frac{4059}{1960}\right]\\
 &\quad-\frac{40025}{1792\sqrt{2}\pi^{5/2}c^{9/4}}E({\scriptstyle{\frac52}},\tau)+O(c^{-5/2})\,,\\
}
where the non-holomorphic Eisenstein series $E(s,\tau,\bar\tau)$ are defined in \eqref{EisensteinExpansion}.

In the $1/c^2$ term for $\lambda^2_{4,0}$, the rational term comes from the contact term ambiguity as fixed from localization in \cite{Chester:2019pvm}, and the coefficient $a$ comes from the one-loop expression in \cite{Aprile:2017bgs,Alday:2017xua,Alday:2021peq}. The calculation that gives the value of $a$ is described in Appendix \ref{1loop}, and can be carried out to any desired precision, e.g.
\es{a}{
a \approx 3.5897946432786394668\,.
}

\section{Numerical bootstrap}
\label{numBoot}

We now combine the numerical bootstrap with the two integrated constraints \eqref{constraint1} to bound CFT data as a function of $\gym$ for many values at $N$. We will throughout set $\theta = 0$, because it has been observed previously that the instanton effects controlled by $\theta$ have only a very small effect on the CFT data of the lowest unprotected scalar in the fundamental domain of $\tau$ \cite{Chester:2021aun}. 

In Section \ref{setup2}, we will describe a slight variation of the approach introduced in \cite{Chester:2021aun} that we find leads to more numerically stable bootstrap problems. In Section \ref{results}, we then present our bootstrap bounds on scaling dimensions and OPE coefficients, and compare with the perturbative results of the previous section.

\subsection{Bootstrap algorithm}
\label{setup2}

Without the integrated constraints \eqref{constraint1}, we could use the approach of \cite{Beem:2016wfs}. Very briefly, we define the functions
\begin{equation}
\begin{split}
	F_{\Delta,\ell}(U,V) &\equiv V^4 G_{\Delta + 4,\ell}(U,V) - U^4 G_{\Delta + 4,\ell}(V,U)\,, \\
	F_\text{short}(U,V) &\equiv U^2 V^4 \cT_\text{short}(U,V) - U^4 V^2 \cT_\text{short}(V, U)\,,
\end{split}
\end{equation}
so that we can write the crossing equation \eqref{CrossAgain} as
\begin{equation}\label{eq:crossing_noics}
	\sum_{\cO_{\Delta,\ell}} \lambda^2_{\Delta,\ell} F_{\Delta,\ell}(U,V) + F_\text{short}(U,V) = 0\,.
\end{equation}
We then expand this equation in derivatives $\partial^m_z \partial^n_{\bar z}$, with $m<n$ and $m+n \in \{1,3,\ldots, \Lambda\}$, at the crossing-symmetric point $z = \bar z = \frac{1}{2}$. If we can find a functional parametrized by coefficients $\alpha_{m,n}$ such that
\begin{equation}\label{eq:bootstrap_problem_noics}
\begin{split}
	\sum_{m,n} \alpha_{m,n} \left(\left.\partial_z^m \partial_{\bar z}^n F_{\Delta,\ell}\right|_{z = \bar z = 1/2}\right) &\ge 0 \qquad \forall \ell = 0,2,\ldots,\quad\Delta \ge \begin{cases} \Delta_* & \ell = 0,\\ \ell + 2 & \ell > 0 \end{cases},\\
	\sum_{m,n} \alpha_{m,n} \left(\left.\partial_z^m \partial_{\bar z}^n F_\text{short}\right|_{z = \bar z = 1/2}\right) &= 1,
\end{split}
\end{equation}
then we have a contradiction with \eqref{eq:crossing_noics}, and so the dimension of the lowest unprotected scalar must satisfy $\Delta_0 < \Delta_*$. A similar algorithm can be used to bound the OPE coefficient of the lowest dimension unprotected scalar.

In order to make the bootstrap sensitive to the value of the coupling $\gym$, we also have to impose the integrated constraints in the form \eqref{IntAgain}. We would thus include additional parameters $\alpha_2$ and $\alpha_4$ in our functional, and search for functionals satisfying
\begin{equation}\label{eq:bootstrap_problem}
\begin{split}
	\alpha_2 I_2\left\lbrack \frac{G_{\Delta+4,\ell}(U,V)}{U^2}\right\rbrack + \alpha_4 I_4\left\lbrack \frac{G_{\Delta+4,\ell}(U,V)}{U^2}\right\rbrack + \sum_{m,n} \alpha_{m,n} \left(\left.\partial_z^m \partial_{\bar z}^n F_{\Delta,\ell}\right|_{z = \bar z = 1/2}\right) &\ge 0,\\
	\alpha_2\left(I_2[\cT_\text{short}] -\cF_2(\tau,\bar\tau)\right) + \alpha_4\left(I_4[\cT_\text{short}] -\cF_4(\tau,\bar\tau)\right)+\sum_{m,n} \alpha_{m,n} \left(\left.\partial_z^m \partial_{\bar z}^n F_\text{short}\right|_{z = \bar z = 1/2}\right) &= 1.
\end{split}
\end{equation}
However, we run into a problem when we compute the integrals of the blocks appearing in \eqref{eq:bootstrap_problem}. We will describe this problem below, a resolution to it used previously in \cite{Chester:2021aun}, and a different resolution that we will employ here.

The integrated constraints were written in \eqref{ints} as integrals over the entire real plane, but the block expansion in \eqref{Gexp} only converges for a subset of that domain. In terms of the radial variables \cite{Hogervorst:2013sma} 
\es{retatozzb}{
U=\frac{16 r^2}{\left(r^2+2 \eta  r+1\right)^2}\,,\qquad V=\frac{\left(r^2-2 \eta  r+1\right)^2}{\left(r^2+2 \eta 
   r+1\right)^2}\,,
}
the block expansion converges within the domain
\es{fundDomain}{
D_1:\qquad r\leq -\sqrt{4 | \eta | +\eta ^2+3}+| \eta | +2\,,\qquad |\eta|\leq1\,.
}
The crossing symmetry group $S_3$ mentioned in Section~\ref{setup} acts by permuting this domain with two other regions
\es{others}{
D_2&:\qquad r\geq -\sqrt{4 | \eta | +\eta ^2+3}+| \eta | +2\,,\qquad\;\;\, 0\leq\eta\leq1\,,\\
 D_3&:\qquad r\geq -\sqrt{4 | \eta | +\eta ^2+3}+| \eta | +2\,,\qquad -1\leq\eta\leq0\,.
}
After mapping back to the $(U,V)$ plane, every point in this plane belongs to exactly one of these three regions. Thus, to apply the integrated constraints to a block, we can restrict the integration region to $D_1$ and multiply by 3 to get 
\es{intsreta}{
I_2\left[\frac{G_{\Delta+4,\ell}}{U^2}\right]&=-3\int_{D_1} dr\,d\eta\frac{\sqrt{1-\eta ^2} \left(r^2-1\right)^2 \left(\left(2-4
   \eta ^2\right) r^2+r^4+1\right)}{128 \pi  r^5}G_{\Delta+4,\ell}(r,\eta)\,,\\
   I_4\left[\frac{G_{\Delta+4,\ell}}{U^2}\right]&=3\int_{D_1} dr\,d\eta \Bigg\lbrack \frac{\sqrt{1-\eta ^2} \left(r^2-1\right)^2 \left(2
   \eta  r-r^2-1\right) \left((4 \eta ^2+10)
   r^2+r^4+1\right)}{4 \pi  r^5 \left(r^2+2 \eta  r+1\right)} \\
   &\qquad\quad \times \bar{D}_{1,1,1,1}G_{\Delta+4,\ell}(r,\eta)\Bigg\rbrack\,.\\
}
Crossing symmetry implies that this is equivalent to \eqref{ints}. For a given $\Delta$ and $\ell$, we can expand the blocks, the integration measures, and $\bar{D}_{1,1,1,1}$ all at small $r$ to some order $p$, perform the integrals over $r$ exactly, and then the remaining integrals over $\eta$ numerically. The error from this expansion scales as $r_\text{max}^p$, where $r_\text{max}=2-\sqrt{3}\approx .268$ is the maximum value of $r$ in $D_1$, and so it converges quite quickly as $p$ is increased.

However, this maximum value of $r$ in $D_1$ occurs at $\eta=0$ where the blocks oscillate in sign between spins $\ell \equiv 0\pmod{4}$ and $\ell\equiv 2\pmod{4}$. The dominant contribution to the integral comes from the vicinity of this point in $D_1$, so it follows that the integrated blocks have the same sign oscillations. This presents a problem for the bootstrap, because for large $\Delta$ the integrated blocks grow as $(4r_\text{max})^\Delta$ while the block derivatives grow as $(4(3-2\sqrt{2}))^\Delta$. For $D_1$, we have $r_\text{max} > 3-2\sqrt{2}$, and so for large $\Delta$ the integrated constraints dominate in \eqref{eq:bootstrap_problem}. The sign oscillations would then force $\alpha_2 = \alpha_4 = 0$ for a functional that is positive at large $\Delta$, and thus the integrated constraints would have no effect.

One way to circumvent this problem, which we used in \cite{Chester:2021aun}, is to consider deformed integration regions $D'(b)$ in which $r_\text{max} = b > 2-\sqrt{3}$ occurs at $\eta = \pm 1$ where the blocks do not exhibit these sign oscillations. The regions $D'(b)$ are equivalent to $D_1$ under crossing symmetry, and thus in the $\Lambda \to \infty$ limit where we recover exact crossing symmetry in the bootstrap, the results should be insensitive to the choice of the integration region. However, at finite $\Lambda$ this is no longer the case, and switching to $D'(b)$ allows $\alpha_2$ and $\alpha_4$ to be nonzero in a positive functional, so that the integrated constraints have an effect on the bootstrap bounds.

In fact, in \cite{Chester:2021aun} we found that the bounds become significantly stronger once we impose the integrated constraints twice, using two integration regions with different values of $b$. In the $\Lambda\to\infty$ limit these two sets of constraints would be redundant, but at finite $\Lambda$ they are not.

This is a satisfying resolution to the problem, in that it allows us to obtain functionals that are positive everywhere and thus obtain rigorous bounds (up to the effects of the various finite truncations we will describe below). However, it leads to another problem, this time one of numerical stability. Since the two sets of integrated constraints are exactly redundant as $\Lambda\to\infty$, they become nearly redundant at large $\Lambda$. Empirically, we found that this leads to serious numerical difficulties at $\Lambda > 39$ (the value that was used in \cite{Chester:2021aun}). In this paper it is crucial for us to reach higher values of $\Lambda$, as we will see in Section \ref{numBoot}, so we have to use a different approach to solve the problem of sign oscillations in the integrated blocks at large $\Delta$.

In this paper, we will avoid the oscillations by only imposing positivity of the functional up to $\Delta = \ell + h_\text{max}$. This allows us to obtain functionals with any integration region, and so for simplicity we will use $D_1$. This means we are guaranteed that functionals with nonzero $\alpha_2$ or $\alpha_4$ will be negative for half of the spins at large $\Delta$, so the clear question is whether our bounds are trustworthy. We have two kinds of evidence that they are.

First, empirically, we have observed that there is a wide range of cutoffs $h_\text{max}$ for which the bounds obtained using the cutoff method are nearly identical to those obtained using the two non-oscillating regions, for values of $\Lambda$ at which we can use either method. We have checked this for the bounds obtained in this paper. Additionally, we used the cutoff method in the \href{https://arxiv.org/abs/2111.07989v1}{first version} of \cite{Chester:2021aun}, and the bounds we obtained were essentially identical to what we obtained in the final version using two non-oscillating regions.

This phenomenon begs for an explanation: why do bounds obtained using functionals that we know not to be positive coincide with the bounds obtained in a fully rigorous approach? Our second piece of evidence lies in the answer to this question: the value of $h_\text{max}$ can be increased with $\Lambda$. That is, although at a fixed $\Lambda$ we can only make $h_\text{max}$ so large before the sign oscillations set in and the bounds weaken to their values without integrated constraints, as $\Lambda$ increases we can make $h_\text{max}$ steadily larger while still retaining the effects of the integrated constraints. In the $\Lambda\to\infty$ limit, we would recover functionals positive for arbitrarily large $\Delta$.

\begin{figure}
\begin{center}
	\includegraphics[width=0.7\linewidth]{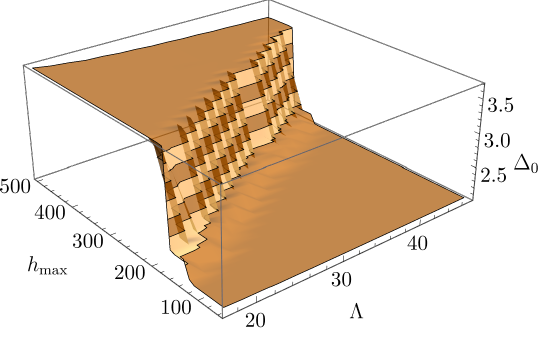}
	\caption{The best bound we can obtain on the scaling dimension of the lowest scalar in the $\grSU(4)$ theory at $\frac{\gym^2}{4\pi} = \frac{1}{100}$ as a function of $\Lambda$ and $h_\text{max}$. For any value of $h_\text{max}$, there exists a $\Lambda$ large enough that we can obtain strong bounds that match the weak coupling expansion as in Figure~\ref{fig:lowN_dimension}.}
	\label{fig:lambda_vs_delta}
	\end{center}
\end{figure}

To see how this works, recall that at any fixed spin $\ell$, the asymptotic scaling of a block derivative can be written as
\begin{equation}
\left.\partial_z^m \partial_{\bar z}^n G_{\ell+h,\ell}\right|_{z=\bar z=1/2} \underset{\tiny h\to\infty}{\sim} h^{m+n}\left(4(3-2\sqrt{2})\right)^h.
\end{equation}
Comparing this with the asymptotic behavior of the integrated blocks, we find that the latter begin to dominate once 
\begin{equation}
	h^\Lambda\left(4(3-2\sqrt{2})\right)^h \approx \left(4(2-\sqrt{3})\right)^h.
\end{equation}
Solving this gives
\begin{equation}
	h_\text{max} \approx -\frac{\Lambda}{C} W_{-1}\left(-\frac{C}{\Lambda}\right)
\end{equation}
where $C = \log\frac{2-\sqrt{3}}{3-2\sqrt{2}} \approx 0.446$ and $W_{-1}$ is the branch of the Lambert W function which gives the real solution we are interested in. For large $\Lambda$, this relation becomes
\begin{equation}\label{eq:lambda_vs_delta}
	h_\text{max} \approx \frac{\Lambda}{C}\left(\log\frac{\Lambda}{C} + \log\left(\log\frac{\Lambda}{C}\right)\right).
\end{equation}

This means that if we want to find functionals positive up to some cutoff $h_\text{max}$, we need to take sufficiently large\footnote{In \cite{Lin:2015wcg}, it is reported that the integrated constraints have no effect on the bounds until a critical $\Lambda$ is reached, after which the bounds suddenly improve. This is likely to also be a result of the sort of behavior seen in Figure~\ref{fig:lambda_vs_delta}, though the detailed interpretation differs because in their case the functionals can be rigorously argued to be positive.} $\Lambda$ to ensure we have enough derivatives to balance the scaling of the integrated constraints at large $h$. According to \eqref{eq:lambda_vs_delta}, the required $\Lambda$ should grow roughly linearly with $h_\text{max}$. Hence, although we do not find strictly positive functionals, we have a method for finding functionals positive up to any specified cutoff. 

Figure~\ref{fig:lambda_vs_delta} shows the best upper bound we can obtain on the scaling dimension of the lowest scalar for the $\grSU(4)$ theory at weak coupling ($\frac{\gym^2}{4\pi} = \frac{1}{100}$) as a function of both $h_\text{max}$ and $\Lambda$. For a fixed $\Lambda$, we see that if $h_\text{max}$ becomes large enough, the bound becomes quite large, in fact reaching the value it would have if we did not include integrated constraints. However, this critical $h_\text{max}$ grows roughly linearly with $\Lambda$, as we expect. Conversely, for any given $h_\text{max}$, we can find some $\Lambda$ for which we will be able to find a functional positive up to that $h_\text{max}$ which imposes a tight bound on $\Delta_0$. This tight bound comes very close to the minimum value it could take according to the weak coupling expansion \eqref{konishiWeak}, as we will see in the following section.

We interpret this behavior as a sign that a fully rigorous proof of our bounds requires the use of \emph{all} of the information from crossing symmetry, or at the very least, more than is contained in any finite truncation of a Taylor series at the crossing-symmetric point. This is consistent with our finding in \cite{Chester:2021aun} that imposing constraints from two integration regions, whose equivalence can only be established using exact crossing symmetry, allows us to find positive functionals. Nevertheless, since the cutoff strategy allows us to obtain functionals positive up to any specified $h_\text{max}$, and a simple model like \eqref{eq:lambda_vs_delta} for the dependence of the required $\Lambda$ on $h_\text{max}$ matches the numerical results, there is good reason to believe that a fully rigorous proof of the bounds obtained in this paper exists.

In the following section, we will show bounds on scaling dimensions and OPE coefficients obtained by converting the continuous set of constraints \eqref{eq:bootstrap_problem} into a finite linear program. This is accomplished by using spins $\ell = \{0,2,\ldots,\ell_\text{max}\}$, cutting off $\Delta \leq \ell + h_\text{max}$ as described above, and using a grid of values of $\Delta$ with finite spacing. This grid, along with further details of the numerical implementation, are given in Appendix \ref{bootApp}.

\subsection{Results}
\label{results}

Using the bootstrap algorithm described above, we can find upper bounds on the scaling dimension $\Delta_0$ and the OPE coefficient $\lambda^2_0$ of the lowest-dimension unprotected operator in $\mathcal{N} = 4$ SYM\@. Here we will present examples of these results, showing where they match perturbative results given in Section \ref{4point} and where they interpolate between weak-coupling and strong-coupling behavior.

\begin{figure}
	\centering
	\includegraphics[width=.8\linewidth]{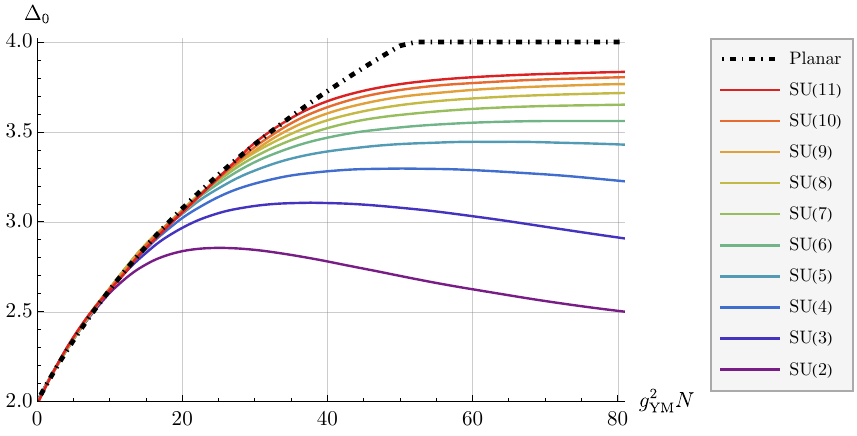}
	\caption{Our bootstrap bounds on the scaling dimension of the lowest unprotected scalar for $2\leq N\leq 11$, extrapolated to $\Lambda\to\infty$ and plotted as a function of the 't Hooft coupling $g_\text{YM}^2 N$. The results in the planar limit obtained from integrability agree with this bound for an increasing large range of the 't Hooft coupling as we increase $N$.}
	\label{fig:integrability}
\end{figure}

In Figure \ref{fig:integrability} we summarize our bounds on $\Delta_0$ as a function of $N$ and $\gym^2 N$, obtained by extrapolating $\Lambda\to\infty$ as described below. Note that some of these curves contain regions with $g_\text{YM}>\sqrt{4\pi}$; the bounds can be obtained in this case by using $\grSL(2,\Z)$ duality to map to a point with $g_\text{YM}<\sqrt{4\pi}$. We also plot the dimension of this operator in the planar limit, like in Figure \ref{fig:integrability_spectrum}. We see that the planar result agrees with the bootstrap bounds for an increasing range of $\gym^2 N$ as we increase $N$.

\begin{figure}
	\centering
	\includegraphics[width=.75\linewidth]{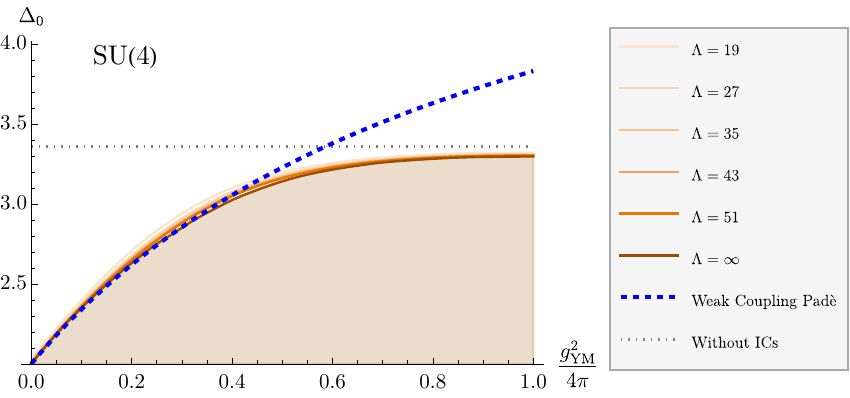}\\[1em]
	\includegraphics[width=.75\linewidth]{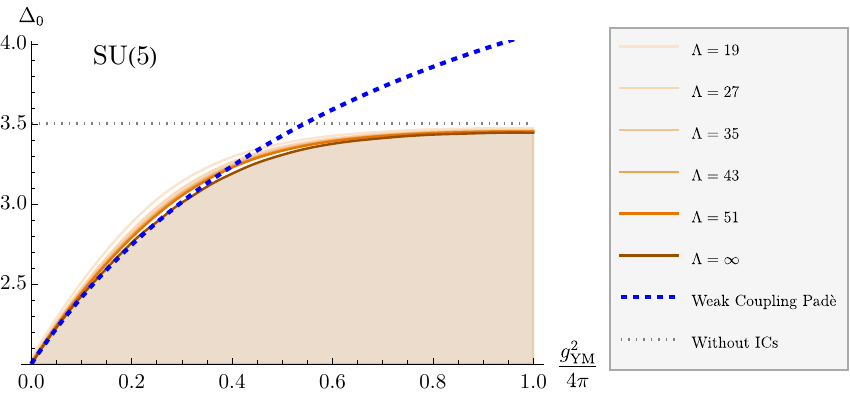}\\[1em]
	\includegraphics[width=.75\linewidth]{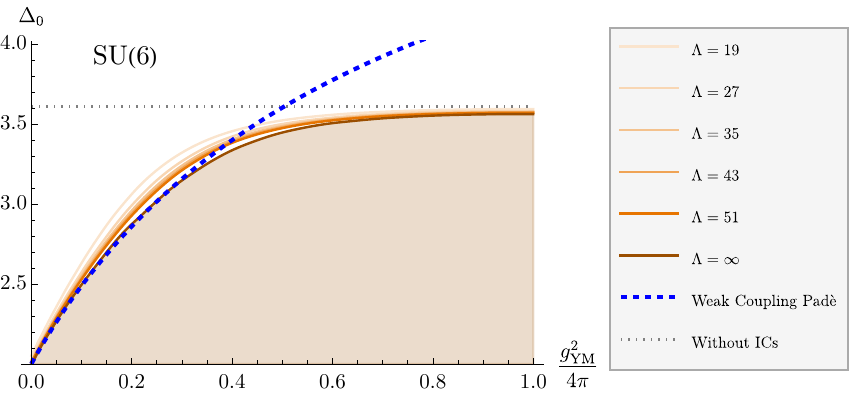}\\[1em]
	\caption{Examples of our bootstrap bounds on the scaling dimension of the lowest unprotected scalar as a function of the Yang-Mills coupling from the free theory to the self-dual point for the $\grSU(4)$, $\grSU(5)$, and $\grSU(6)$ theories. In each case we compare with a (2,2) Pad\'e approximant to the weak coupling expansion \eqref{konishiWeak}. This agrees excellently with our bootstrap bounds for small $g_\text{YM}$ extrapolated to $\Lambda\to\infty$, although the convergence in $\Lambda$ becomes slower as we increase $N$. The $g_\text{YM}$-independent bounds we would obtain without integrated constraints are shown in the dotted gray lines.}
	\label{fig:lowN_dimension}
\end{figure}

In Figure~\ref{fig:lowN_dimension}, we show in more detail the upper bound on $\Delta_0$ as a function of $\frac{\gym^2}{4\pi}$ for $N = 4,5,$ and $6$, from the free theory point to the self-dual point $\gym^2 = 4\pi$ (see Appendix \ref{bootApp} for similar plots at $2\leq N\leq 11$). We plot the upper bounds obtained at several finite values of $\Lambda$. The convergence in $\Lambda$ significantly worsens as we increase $N$. Thus, while in \cite{Chester:2021aun} we could obtain well-converged bounds for $N = 2,3$ even at finite $\Lambda$, here we must extrapolate to $\Lambda\to\infty$ in order to obtain a good estimate of $\Delta_0$. See Appendix \ref{bootApp} for an example of this extrapolation.

We see in Figure~\ref{fig:lowN_dimension} that a $(2,2)$ Pad\'e approximant to the weak coupling expansion \eqref{konishiWeak} and the results from integrability both match our bounds very well near the weak-coupling limit. We have found this agreement for every value of $N$ we have run, although since the convergence in $\Lambda$ is slower at larger $N$, we are limited in how high in $N$ we can go before the extrapolation to $\Lambda\to\infty$ ceases to be trustworthy.

\begin{figure}
	\centering
	\includegraphics[width=.75\linewidth]{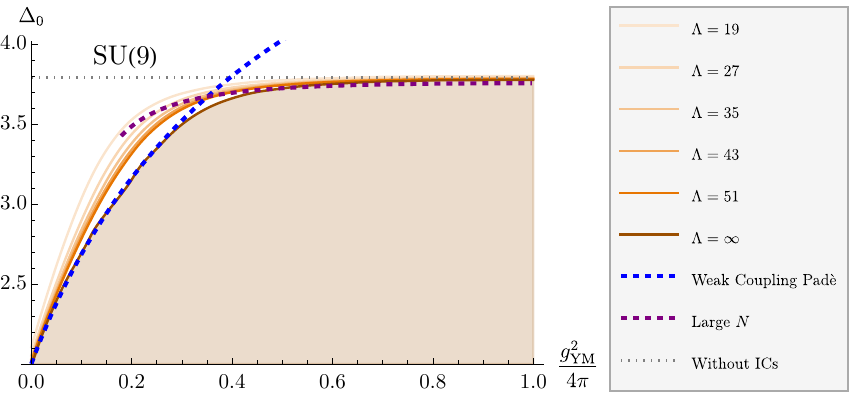}\\[1em]
	\includegraphics[width=.75\linewidth]{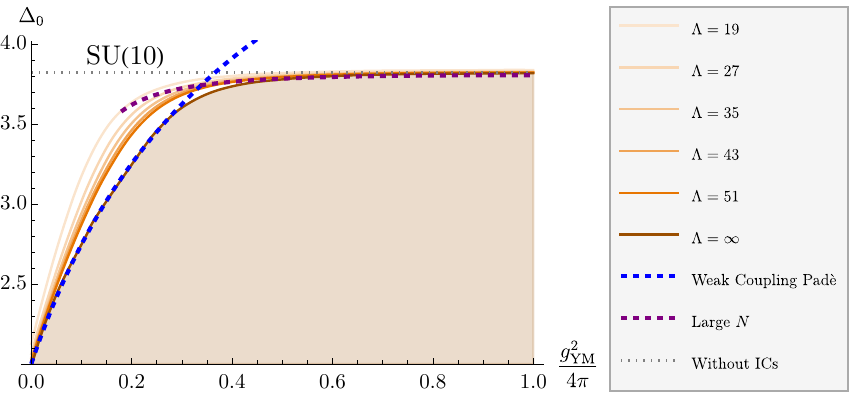}\\[1em]
	\includegraphics[width=.75\linewidth]{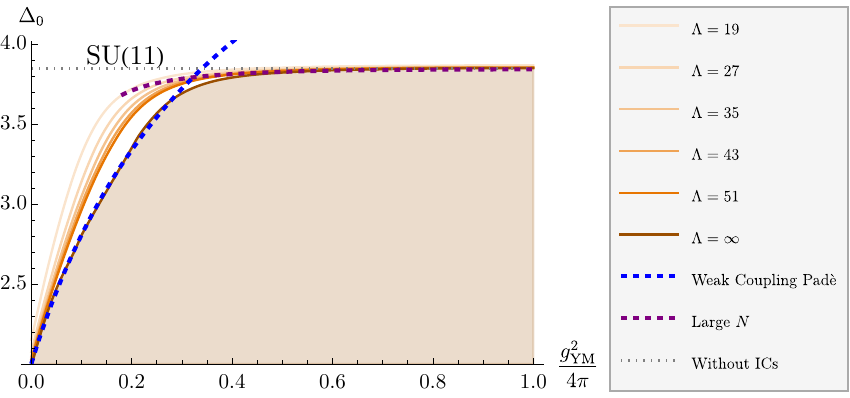}\\[1em]
	\caption{Our bootstrap bounds on the dimension of the lowest scalar as a function of the Yang-Mills coupling from the free theory to the self-dual point for the $\grSU(9)$, $\grSU(10)$, and $\grSU(11)$ theories. We compare with a (2,2) Pad\'e approximant to the weak coupling expansion \eqref{konishiWeak} for the scaling dimension of the twist-two scalar. In addition, we plot the large $N$, finite $g_\text{YM}$ expansion \eqref{largeN} for the twist-four double-trace operator. Our bound on the lowest-dimension scalar interpolates between these two expansions in a way suggestive of a level-repulsion between the twist-two and twist-four operators where their dimensions become close. The $g_\text{YM}$-independent bounds we would obtain without integrated constraints are shown in the dotted gray lines.} 
	\label{fig:highN_dimension}
\end{figure}

Nevertheless, there is a range of $N$ for which we still have sufficiently good convergence to extrapolate $\Lambda\to\infty$, but where the large $N$ results \eqref{largeN} match our bootstrap bounds well at strong coupling, where the lowest-dimension unprotected scalar operator is double-trace. In Figure~\ref{fig:highN_dimension}, we plot our bounds for $N = 9,10,$ and $11$, and compare with both the weak-coupling and the large $N$ results. The bootstrap bound at $\Lambda\to\infty$ is suggestive of a level repulsion between the single-trace and double-trace operators. As $N\to\infty$ these levels sharply cross as in Figure \ref{fig:integrability_spectrum}, and indeed the level repulsion region is shrinking as $N$ increases.

\begin{figure}
	\centering
	\includegraphics[width=.8\linewidth]{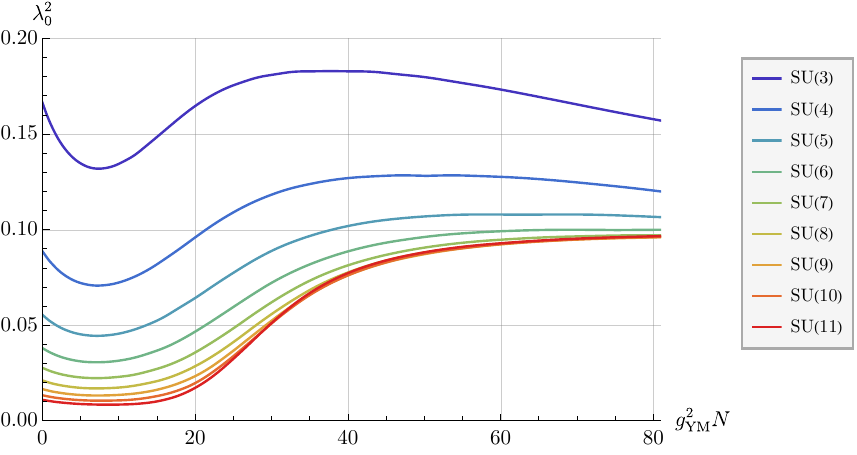}
	\caption{Our bootstrap bounds on the OPE coefficient of the lowest unprotected scalar for $3\leq N\leq 11$, extrapolated to $\Lambda\to\infty$ and plotted as a function of the 't Hooft coupling $g_\text{YM}^2 N$. In this case there are no integrability results with which to compare.}
	\label{fig:ope_lambda}
\end{figure}

In addition to the scaling dimension of the lowest operator, we can use the bootstrap to extremize its OPE coefficient in the OPE of $S\times S$. We thereby obtain an upper bound on this quantity. In Figure \ref{fig:ope_lambda}, we plot this bound for $2\leq N\leq 11$ as a function of $\gym^2 N$. In this case there is no known result from integrability to compare with.

\begin{figure}
	\centering
	\includegraphics[width=.75\linewidth]{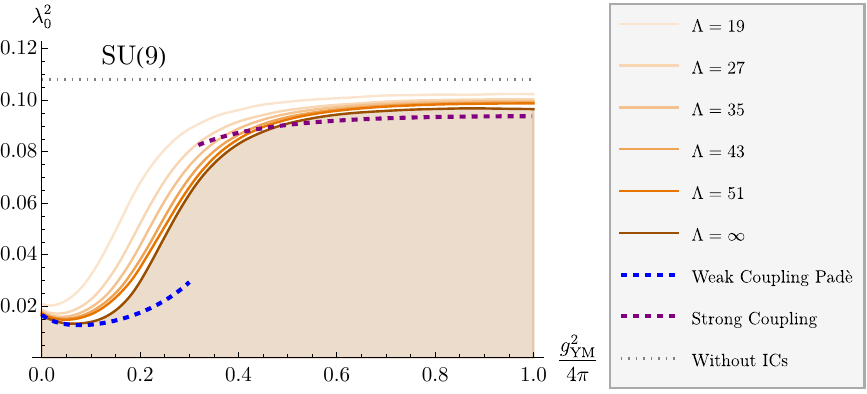}\\[1em]
	\includegraphics[width=.75\linewidth]{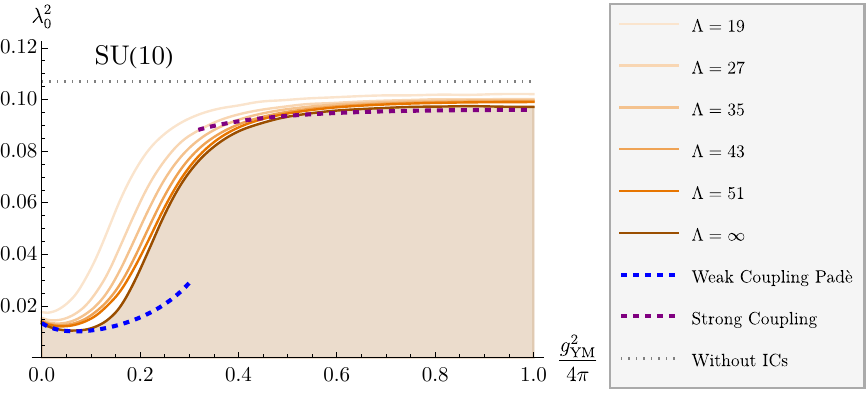}\\[1em]
	\includegraphics[width=.75\linewidth]{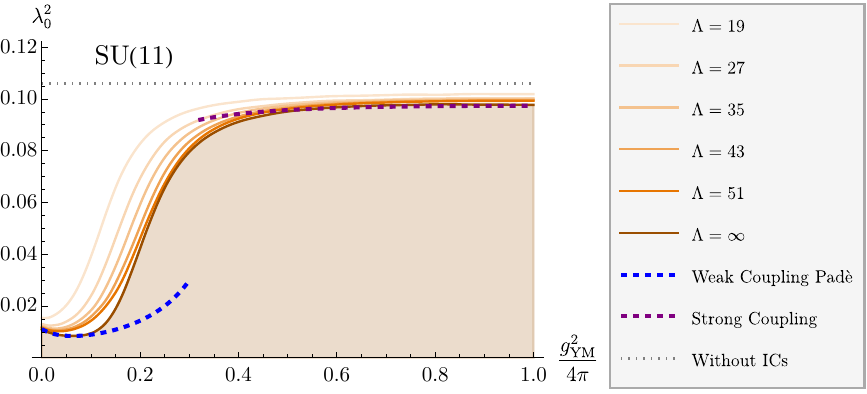}\\[1em]
	\caption{Examples of our bootstrap bounds on the OPE coefficient of the lowest scalar as a function of the Yang-Mills coupling from the free theory to the self-dual point, for the $\grSU(9)$, $\grSU(10)$, and $\grSU(11)$ theories. We compare with a (2,2) Pad\'e approximant to the weak coupling expansion \eqref{konishiWeak} for the OPE coefficient of the twist-two scalar, along with the large $N$, finite $g_\text{YM}$ expansion \eqref{largeN} for the OPE coefficient of the twist-four double-trace operator. Our bounds exhibit increasingly sharp jumps between these two limits in a way suggestive of a level-repulsion between the twist-two and twist-four operators where their dimensions become close.} 
	\label{fig:ope}
\end{figure}

\begin{figure}
	\centering
	\includegraphics[width=\linewidth]{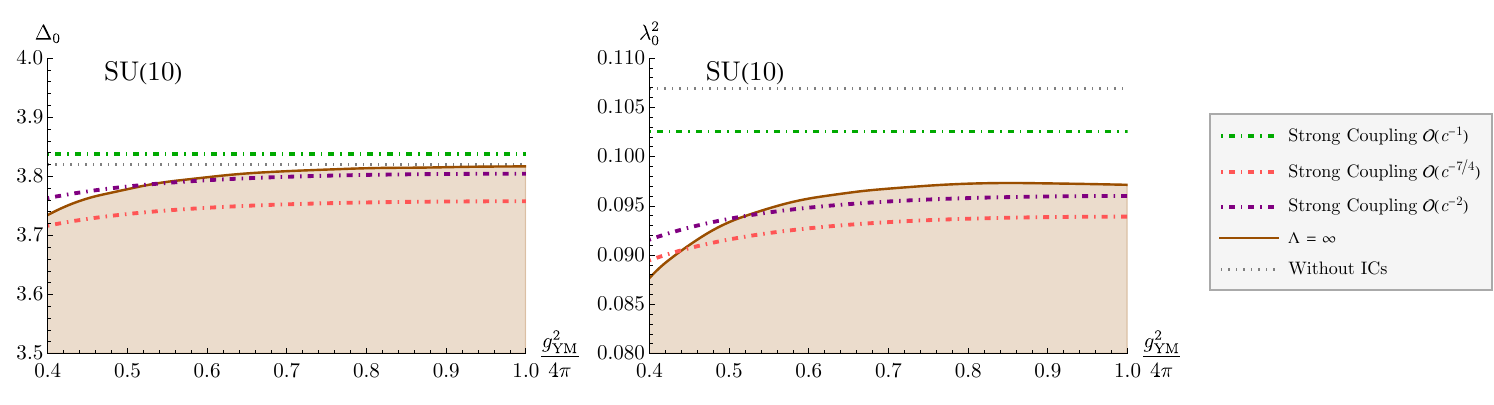}
	\caption{The bootstrap bounds as $\Lambda\to\infty$ in the strong-coupling region for both the scaling dimension and the OPE coefficient of the lowest-dimension operator for $\grSU(10)$, compared with the large $N$ expansion \eqref{largeN} for these quantities. As we include more terms in \eqref{largeN}, the agreement with our bounds improves.}
	\label{fig:strong_coupling}
\end{figure}

We find a good match between our OPE coefficient bounds and a (2,2) Pad\'e approximant to the weak-coupling expansion for the OPE coefficient of the Konishi operator \eqref{konishiWeak}. In addition, for sufficiently large $N$ we can compare the behavior at strong coupling with the large $N$ result for the double-trace operator.

In Figure~\ref{fig:ope} we plot our bounds on the OPE coefficient of the lowest-dimension operator for $\grSU(9)$, $\grSU(10)$, and $\grSU(11)$  (see Appendix \ref{bootApp} for similar plots at $2\leq N\leq 11$). After extrapolating to $\Lambda\to\infty$, we see good agreement with the weak-coupling expansion, and also good agreement with the large $N$ result in the strong-coupling region. Between them, the bounds exhibit increasingly sharp jumps between the values in the two limits, which is also suggestive of a level repulsion as the two operators cross.

Finally, we can take a closer look at the agreement between the large $N$ expansion \eqref{largeN} and our bounds for large $g_\text{YM}^2 N$. It is interesting to check whether the agreement improves as we include more terms from the expansion. In Figure~\ref{fig:strong_coupling}, we zoom in on the strong-coupling region for $\grSU(10)$, and compare the large $N$ expansion of both the scaling dimension and OPE coefficient of the twist-four double-trace operator with our bootstrap bounds at order $\frac{1}{c}$, $\frac{1}{c^{7/4}}$, and $\frac{1}{c^2}$ in the expansion. In both cases, we see that the agreement is improving as we include more terms. However, we found that including the $\frac{1}{c^{9/4}}$ term significantly worsens the agreement, which may be due to the asymptotic nature of the expansion.

\section{Discussion}
\label{disc}

In this paper, we computed bounds on the scaling dimension and OPE coefficient of the lowest-dimension unprotected scalar operator in $\grSU(N)$ $\mathcal{N}=4$ SYM for a wide range of $N$ and $g_\text{YM}$ (at $\theta=0$), thus extending the analysis of \cite{Chester:2021aun} for $N=2$ and $3$ to larger $N$. We used a modified integration region for the integrated constraints which allowed us to compute bounds at larger bootstrap precision $\Lambda$, which was necessary to address the slower convergence at larger $N$. As in \cite{Chester:2021aun}, we showed that weak coupling results saturate these bounds in the appropriate regime. Since we now have access to larger values of $N$, we were able to show that, at large $N$, our bounds interpolate from the CFT data of the Konishi operator at weak coupling to that of the lowest double-trace operator at strong coupling, whose CFT data is computed from string theory on $AdS_5\times S^5$.  In order to obtain a good match at strong coupling, we needed to include several orders in the strong coupling expansion in our comparison to the numerical bounds.  Our bounds also describe an intermediate regime inaccessible to either perturbative expansion where level repulsion occurs as the lowest dimension operator changes from the single- to the double-trace operator. We also matched our scaling dimension bounds to planar integrability results at weak coupling, which show that non-planar corrections are small in this regime.

Looking ahead, our main goal is to further strengthen the bootstrap so that we can accurately estimate the value of the subleading scalar operator or higher spin operators. Since perturbative methods suggest that there are only two relevant unprotected operators, the Konishi and the double-trace scalar operator with dimension 4 in the planar limit, imposing the existence of just two relevant operators should allow us to compute precise islands for CFT data as a function of $\tau$, as long as we are sensitive to the next lowest operator. To achieve this sensitivity, we could impose additional integrated constraints from correlators of different half-BPS operators \cite{Binder:2019jwn}, whose localization input was recently computed at finite $N$ and $\tau$ in \cite{Brown:2023cpz,Brown:2023why,Paul:2023rka,Paul:2022piq}, or from possibly independent constraints coming from derivatives of the squashed sphere free energy. We could also consider mixed correlators with the relevant long operators, as mixing all relevant operators was necessary to find islands in previous bootstrap studies in non-supersymmetric CFTs \cite{Kos:2014bka,Kos:2015mba}.

One question about string theory that we can address with improved precision even for low-lying CFT data is about higher-derivative corrections to the type IIB effective action. The first few corrections to supergravity, namely the $R^4$ and $D^4R^4$ terms in the string effective action, were already computed analytically in \cite{Binder:2019jwn,Chester:2019jas,Chester:2020vyz,Chester:2020dja,Chester:2019pvm}  using as input the localization constraints described in Section \ref{4point}. It may be possible to similarly compute the last protected correction $D^6R^4$ using new constraints from the squashed sphere, but higher order terms starting with $D^8R^4$ are unprotected and so probably cannot be fixed analytically. In this paper, we showed that our numerical bootstrap bounds are sensitive to at least the first few correction terms, but we did not perform a systematic analysis. We leave such an analysis for future work. We hope that with significantly increased bootstrap precision, we would be able to extract the coefficients of these terms in the string effective action from numerical bootstrap results.

Regarding the comparison to integrability, our work provides predictions for CFT data beyond the planar limit. Recently, a method of going beyond the planar limit using integrability has also been discussed \cite{Bargheer:2017nne}, but has so far been implemented only at weak coupling \cite{Bargheer:2018jvq}. A proposal was also made for how to compute OPE coefficients using integrability in the planar limit \cite{Basso:2015zoa}, but so far strong coupling results have only been obtained in the large R-charge limit. Another approach for computing planar OPE coefficients is to input the planar scaling dimensions, as computed from integrability, into the numerical bootstrap. While this method has worked well for the defect CFT defined on a Wilson line in SYM \cite{Cavaglia:2022qpg,Cavaglia:2021bnz,Cavaglia:2022yvv}, it seems more challenging for the full SYM theory \cite{Caron-Huot:2022sdy}. Planar OPE coefficients at strong coupling for single-trsace operators can also be computed using the relation to a recently conjectured form for the AdS Virasoro-Shapiro amplitude \cite{Alday:2023mvu,Alday:2023jdk,Alday:2023flc,Alday:2022xwz,Alday:2022uxp}. Our new data can provide a benchmark for all these various methods.

The method of combining bootstrap with localization can be applied to a host of other theories in various dimensions. So far, the only other application has been to 4d $\mathcal{N}=2$ $\grSU(2)$ conformal SQCD \cite{Chester:2022sqb}, where bounds on unprotected scaling dimensions were also computed as a function of the complexified Yang-Mills coupling. It would be interesting to generalize this to higher rank conformal SQCD, or to the $\grUSp(2N)$ theory considered in \cite{Behan:2023fqq}, which is one of the simplest models of holography for open string theory or F-theory. It would also be nice to strengthen the 3d ABJM bootstrap of \cite{Chester:2014fya,Chester:2014mea,Agmon:2017xes,Agmon:2019imm,Binder:2020ckj} using integrated constraints, which could allow us to read off the $D^8R^4$ correction to the holographic correlator in M-theory, as initiated in \cite{Chester:2018aca,Alday:2021ymb,Alday:2022rly}. Lastly, the 5d $\mathcal{N}=1$ bootstrap \cite{Chang:2017cdx} could also be improved using integrated constraints.

\section*{Acknowledgments} 

We thank Luis Fernando Alday, Tobias Hansen, and Julius Julius for useful discussions. SSP and RD are supported in part by the US NSF under Grant No.~2111977\@.   SMC is supported by the Royal Society under the grant URF\textbackslash R1\textbackslash 221310. RD was also supported in part by an NSF GRFP.

\appendix

\section{Contribution of short multiplets to $\cT(U, V)$}
\label{app:tshort}

The function $\cT(U,V)$ that appears in the stress tensor four-point function can be decomposed into contributions from the long and short representations of the superconformal algebra, as in \eqref{TExp}. The scaling dimensions and OPE coefficients of the long multiplets are unprotected, and these are the quantities we bootstrap in this paper. The contribution from the short multiplets is fixed by superconformal symmetry, and was determined analytically in \cite{Beem:2016wfs}. The contribution $\cT_\text{short}(U,V)$ is $\cT^{(0)}_\text{short}(U,V) + c^{-1} \cT^{(1)}_\text{short}(U,V)$, where
\es{Fshort}{
 \cT^{(0)}_\text{short}=&-1 - \frac{1}{(1-z)^2(1-\bar z)^2}+\frac{24
   \log (1-z) \log (1-\bar z)}{z^2 \bar z^2}\\
   & + \frac{6 \left(-2 z \bar z \left(z^2+z \bar z+\bar z^2-4\right)+(z+\bar z) \left(z^2
   \bar z^2+z^2+\bar z^2-6\right)+4\right)}{(1-z)^2 z (1-\bar z)^2 \bar z}\\
   &+\frac{2 \left(z \left(2
   \bar z^4-\bar z^3+4 \bar z^2-18 \bar z+12\right)-3 \left(\bar z^4-6
   \bar z^2+4 \bar z\right)\right) \log (1-z)}{z^2 (1-\bar z)^2 \bar z
   (z-\bar z)}\\
   &+\frac{2 \left(3 \left(z^4-6 z^2+4 z\right)-\left(2 z^4-z^3+4 z^2-18
   z+12\right) \bar z\right) \log (1-\bar z)}{(1-z)^2 z \bar z^2 (z-\bar z)}\,,\\
 \cT^{(1)}_\text{short}=& -\frac{1}{(1-z)(1-\bar z)} + \frac{36 \log (1-z) \log (1-\bar z)}{z^2 \bar z^2}-\frac{2 \left(\frac{9
   \bar z-18}{z^2 \bar z}+\frac{4}{z-\bar z}-\frac{4}{z}\right) \log
   (1-z)}{1-\bar z}\\
   &-\frac{2 \left(\frac{9 z-18}{z
   \bar z^2}-\frac{4}{z-\bar z}-\frac{4}{\bar z}\right) \log
   (1-\bar z)}{1-z}+\frac{18 \left(\frac{1}{(1-z) (1-\bar z)}+1\right)}{z \bar z}\,.
 }
 
This function enters into the bootstrap through its localization integrals (see \eqref{IntAgain} and \eqref{eq:bootstrap_problem}). These can be computed numerically to any desired precision. They are given by
\es{exactInts}{
&I_2[\cT_\text{short}] \approx 0.0462845727+\frac{0.3895281312 }c\,,\\
&I_4[\cT_\text{short}] \approx 5.60637758+\frac{50.86596767 }c\,.
}

\section{One-loop OPE coefficient}
\label{1loop}

Our goal is to extract the OPE coefficient of the scalar long block at one-loop, i.e. $\cT^{R|R}$, which multiplies $1/c^2$ in the large $c$ expansion of $\cT$.\footnote{See \cite{Zhou:2017zaw,Chester:2018lbz} for similar calculations in $\mathcal{N}=8$ SCFTs.} The one-loop amplitude is most conveniently expressed using the Mellin transform
 \es{MellinDef}{
  \cT^{R|R}(U,V)
   = \int_{-i \infty}^{i \infty} \frac{ds\, dt}{(4 \pi i)^2} U^{\frac s2} V^{\frac t2 - 2}
    \Gamma \left[2 - \frac s2 \right]^2 \Gamma \left[2 - \frac t2 \right]^2 \Gamma \left[2 - \frac u2 \right]^2
    M^{R|R}(s, t) \,,
 } 
where $u \equiv 4 - s - t$ and the contours include poles on just one side of the Gamma functions. The one-loop amplitude in Mellin space was written as a double sum in \cite{Alday:2018kkw}:
\es{MellinRR}{
&M^{R|R}(s,t) = \sum_{m,n=2}^\infty\Bigg[ \frac{c_{mn}}{(s-2m)(t-2n)} + \frac{c_{mn}}{(t-2m)(u-2n)}+ \frac{c_{mn}}{(u-2m)(s-2n)}-b_{mn}\Bigg]+C
}
where the coefficients $c_{mn} = c_{nm}$ are
\begin{align}
c^{SU(N)}_{mn} ={}& 
\frac{(m-1)^2 m^2}{5 (m+n-1)}
+\frac{2(m-1)^2 \left(3 m^2-6 m+8\right)}{5(m+n-2)}
-\frac{9 m^4-54 m^3+123 m^2-126 m+44}{5 (m+n-3)}\nonumber\\
&-\frac{4 \left(m^2-4 m+9\right) (m-2)^2}{5 (m+n-4)}
+\frac{6 (m-3)^2 (m-2)^2}{5 (m+n-5)}\,.
\label{SUNcmn}
\end{align}
These coefficients diverge in the large $m \sim n$ limit, so in \cite{Chester:2019pvm} the $b_{mn}$ were chosen to regulate the sum in this limit, while the constant $C$ was then chosen so that the Mellin amplitude matches the position space correlator of \cite{Aprile:2017bgs}, which gave
\es{bC}{
b_{mn} = \frac{9mn}{2(m+n)^3}\,, \qquad
C = -\frac{39}{16} - \frac{13}{8} \pi^2 + 9 \zeta(3)\,.
}
The regularized double sum
\es{boxads5}{
&\Phi(s,t)= \sum_{m,n=2}^\infty \Big[\frac{c_{mn}}{(s-2m)(t-2n)}-\frac{3 m n}{2 (m+n)^3}+\frac{3 m t-4 m+3 n s-4 n}{4 (m+n)^3}\Big]
}
was resummed in \cite{Alday:2021peq} (note that the last term vanishes in the symmetric combination that enters the Mellin amplitude \eqref{MellinRR}). The result is
\begin{align}
\Phi(s,t) ={}& R_0(s,t) \left( \psi ^{(1)}\left(2-\frac{s}{2}\right)+ \psi ^{(1)}\left(2-\frac{t}{2}\right)- \left( \psi ^{(0)}\left(2-\frac{s}{2}\right)-\psi ^{(0)}\left(2-\frac{t}{2}\right) \right)^2 \right) \nonumber\\
& + R_1(s,t) \psi ^{(0)}\left(2-\frac{s}{2}\right)+R_1(t,s) \psi ^{(0)}\left(2-\frac{t}{2}\right)+R_2(s,t)\,,
\label{AdSbox}
\end{align}
where $\psi^{(n)}(x)$ is the $n$th derivative of the Digamma function, and we define the rational functions
\es{R0}{
R_0(s,t) =& \frac{P_0(s,t)}{40(s+t-10) (s+t-8) (s+t-6) (s+t-4) (s+t-2)}\,,\\
R_1(s,t) =& \frac{P_1(s,t)}{40 (s+t-10) (s+t-8) (s+t-4) (s+t-2)}\,,\\
R_2(s,t) =& \frac{P_2(s,t)}{240 (s+t-10) (s+t-8) (s+t-6) (s+t-4) (s+t-2)}-3\zeta(3)
}
in terms of the polynomials
\es{Ps}{
P_0(s,t)={}&15 s^4 t^2+30 s^3 t^3-360 s^3 t^2+15 s^2 t^4-360 s^2 t^3+2304 s^2 t^2-70 s^4 t+1096 s^3 t\\
&-5048 s^2 t +88 s^4-1024 s^3+3552 s^2-70 s t^4+1096 s t^3-5048 s t^2+8640 s t\\
&-4736 s+88 t^4-1024 t^3+3552 t^2-4736 t+2048\,,\\
P_1(s,t)={}&-105 s^3 t^2-45 s^2 t^3+1200 s^2 t^2-75 s^4 t+1350 s^3 t-7720 s^2 t-15 s^5+400 s^4\\
&-3532 s^3 +13008 s^2+250 s t^3-3980 s t^2+17200 s t-21248 s-368 t^3\\
&+4192 t^2-13312 t+12800\,,\\
P_2(s,t)={}&-45 s^4 t^2-630 s^3 t^2-45 s^2 t^4-630 s^2 t^3+19404 s^2 t^2-72 s^5 t+585 s^4 t \\
&+8640 s^3 t-125604 s^2 t-27 s^6+477 s^5-1062 s^4-28908 s^3+239688 s^2-72 s t^5+585 s t^4 \\
&+8640 s t^3-125604 s t^2+520848 s t-683424 s-27 t^6+477 t^5-1062 t^4-28908 t^3\\
&+239688 t^2-683424 t+642816 
+\pi^2 \left(45 s^4 t^2+520 s^3 t^2+45 s^2 t^4+520 s^2 t^3-14544 s^2 t^2\right.\\
&+54 s^5 t-400 s^4 t 
 -7056 s^3 t+89808 s^2 t+9 s^6-164 s^5-648 s^4+25984 s^3-172656 s^2\\
&+54 s t^5-400 s t^4 
-7056 s t^3+89808 s t^2-354528 s t+458560 s+9 t^6-164 t^5-648 t^4\\
&\left.+25984 t^3-172656 t^2 
+458560 t-419328 \right)\,.
}

To extract the scalar OPE coefficient, we consider the block expansion \eqref{Gexp}, and take the limit $U \to 0$ while setting $V = 1$. In this limit, $U^{-2}G_{8, 0}(U, V) \approx U^2$, so we must have
 \es{S15Approx}{
  \cT^{R|R}(U, 1)  = {\lambda_{4, 0}^2} U^2 + \cdots \,.
 }
 at this order in $1/c$ of the correlator.\footnote{Note that if we were to consider higher spin operators, then we would need to subtract other terms that appear $1/c^2$ as discussed in \cite{Alday:2021peq}. At order $1/c$, lower orders in $U$ terms would also appear.} Thus, in order to extract the contribution to $\lambda_{4, 0}^2$ from $M^{R|R}(s,t) $, which we denoted by $a$ in \eqref{largeN}, all we need to do is extract the coefficient of $U^2$ in the small $U$ expansion of the Mellin transform \eqref{MellinDef} of $M^{R|R}(s,t) $ with $V=1$. This is given by the integral 
 \es{lam84}{
a
   &=    -\int_{-i \infty}^{ i \infty} \frac{dt}{8\pi i}\,  
\Gamma \left[2 - \frac t2 \right]^2  \text{Res} \Bigg[\Gamma \left[ \frac{s+t}{2} \right]^2\Gamma \left[2 - \frac s2 \right]^2 M^{R|R}(s,t)\Bigg]_{s=4}\,,
 }
with the $t$ contour obeying $0< \Re t < 4$, and the minus sign is because we closed the $s$ contour to the right. This integral can be performed numerically to any precision, which yields \eqref{a}.

\section{Numerical bootstrap details}
\label{bootApp}

The bootstrap calculations described in Section \ref{numBoot} reduce to evaluating the feasibility of linear programs, in which the variables parametrize a linear functional and the constraints correspond to spins and scaling dimensions at which we impose positivity of the functional, following the setup \cite{Rattazzi:2008pe}. To find the extremal value of the OPE coefficient, we then evaluate the optimal value of a similar linear program.     

We used two linear program solvers, Gurobi and SDPB \cite{gurobi,Simmons-Duffin:2015qma}. Gurobi is a machine-precision solver, and so at low values of $\Lambda$ where this suffices (in practice, $\Lambda \leq 19$), it is faster. However, at larger $\Lambda$ where machine precision does not suffice, we use SDPB with 512 bits of precision. 

When using SDPB, we find that it is important to carefully choose the set of positivity constraints we impose. Although ideally we would use a very fine grid of scaling dimensions for each spin $0, 2, \ldots, \ell_\text{max}$, in practice this causes SDPB to stall, with the step sizes going to zero. Thus, in practice, we need to use a coarser grid in order for the solver to converge. For each spin $\ell$, we impose positivity at the following values of $\Delta$:
\begin{equation}
	\{\ell + 2, \ell + 2.04, \ldots, \ell + 4, \ell + 4.1, \ldots, \ell + 6, \ell + 6.2, \ldots, \ell + 10, \ell + 11, \ldots, \ell + h_\text{max}\}.
\end{equation}
The other parameters we use are standard, and are described in Table \ref{tab:numerical_parameters}.

\begin{figure}
	\centering
	\includegraphics[width=.7\linewidth]{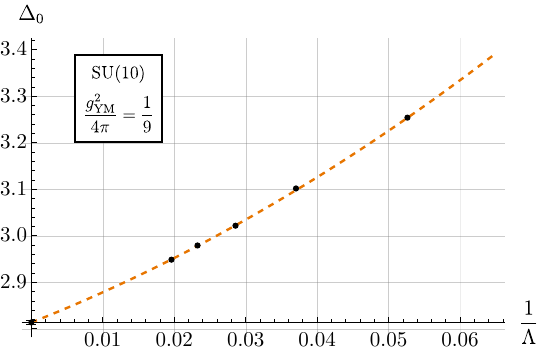}
	\caption{The extrapolation of the upper bound on the lowest scaling dimension for the $\grSU(10)$ theory at $\frac{g_\text{YM}^2}{4\pi} = \frac{1}{9}$. We fit a quadratic polynomial to the bounds at finite $\Lambda$, with $19\leq \Lambda\leq 51$, and then use that polynomial to extrapolate towards $\Lambda\to\infty$. This fit suggests that to obtain a bound within 1\% of this extrapolated value directly from the bootstrap, we would need $\Lambda > 200$, which is very far from being numerically feasible. Thus, the extrapolation procedure is necessary for the comparisons we make in this paper.}
	\label{fig:extrapolation}
\end{figure}

\begingroup
\renewcommand{\arraystretch}{1.1}
\begin{table}
	\centering
	\begin{tabular}{l|p{4cm}p{4cm}}
		$\Lambda$ & $\leq 19$ & $> 19$ \\
		\hline
		Solver & Gurobi & SDPB \\
		\hline
		$\ell_\text{max}$ & 25 & 60 \\
		$h_\text{max}$ & 40 & 60 \\
		\hline
		\texttt{NumericFocus} & 3 & -- \\
		\texttt{DualReductions} & 0 & -- \\
		\hline
		\texttt{precision} & machine ($\sim 53$) & 512 \\
		\texttt{dualityGapThreshold} & -- & $10^{-30}$ \\
		\texttt{primalErrorThreshold} & -- & $10^{-30}$ \\
		\texttt{dualErrorThreshold} & -- & $10^{-30}$ \\
		\texttt{initialMatrixScalePrimal} & -- & $10^{20}$ \\
		\texttt{initialMatrixScaleDual} & -- & $10^{20}$ \\
		\texttt{feasibleCenteringParameter} & -- & $0.1$ \\		
		\texttt{infeasibleCenteringParameter} & -- & $0.3$ \\
		\texttt{stepLengthReduction} & -- & $0.7$ \\
		\texttt{maxComplementarity} & -- & $10^{100}$ \\
	\end{tabular}
	\caption{Parameters used in Gurobi and SDPB for the plots in this paper. We work at up to $\Lambda = 51$, and extrapolate to $\Lambda\to\infty$. Note that the values of $\ell_\text{max}$ and $h_\text{max}$ listed here are the maximum that we use, but lower values would suffice for $\Lambda < 51$.}
	\label{tab:numerical_parameters}
\end{table}
\endgroup

\begin{figure}
	\centering
	\includegraphics[height=.9\textheight]{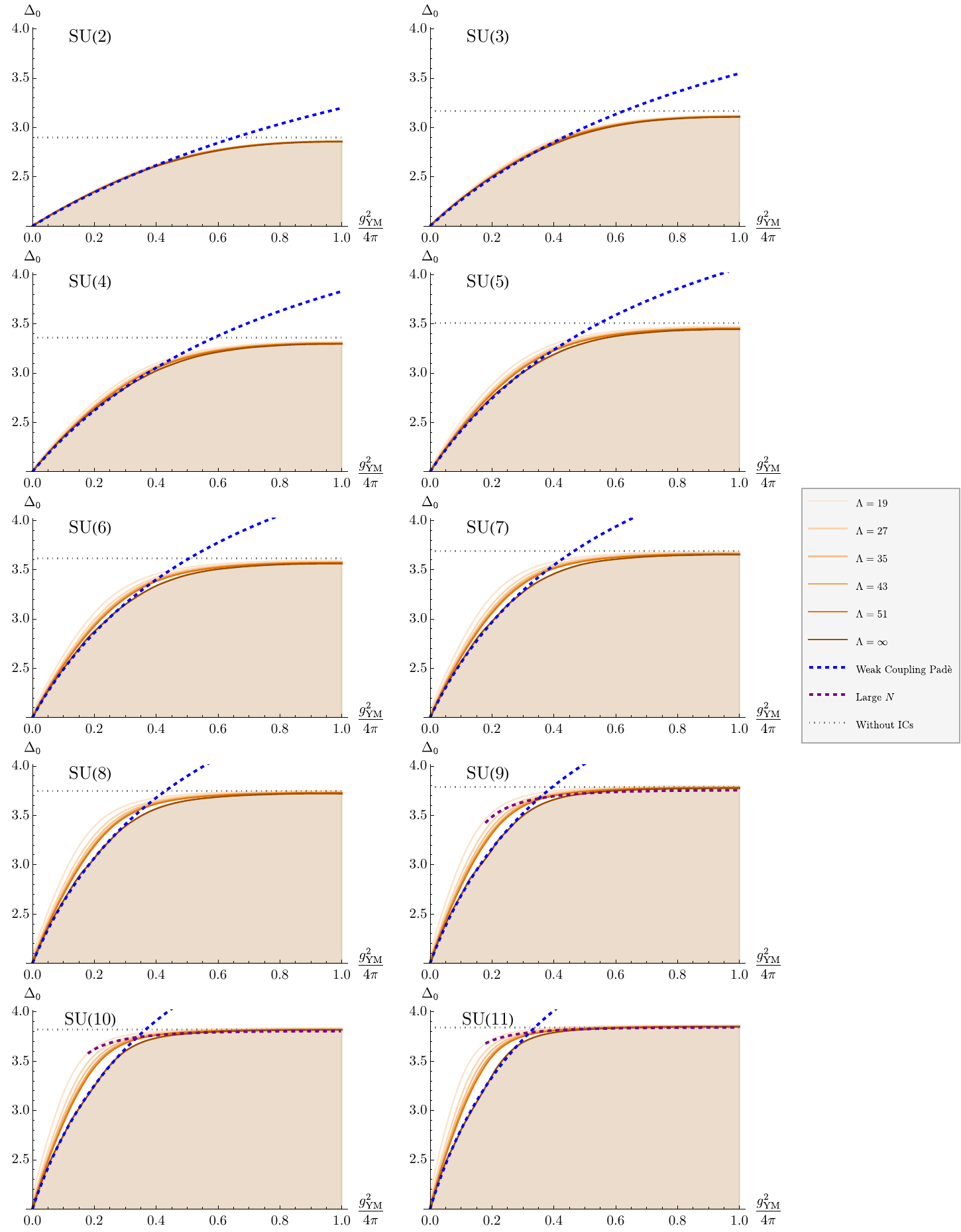}
	\caption{Bootstrap bounds on the scaling dimension of the lowest unprotected scalar operator for $2\leq N\leq 11$, obtained as in Figures~ \ref{fig:lowN_dimension} and \ref{fig:highN_dimension}. For $N\geq 9$, we compare with the large $N$ expansion as in Figure~\ref{fig:highN_dimension}. For $N<9$, this expansion is not sufficiently converged to be of use.}
	\label{fig:dimension_all}
\end{figure}

\begin{figure}
	\centering
	\includegraphics[height=.9\textheight]{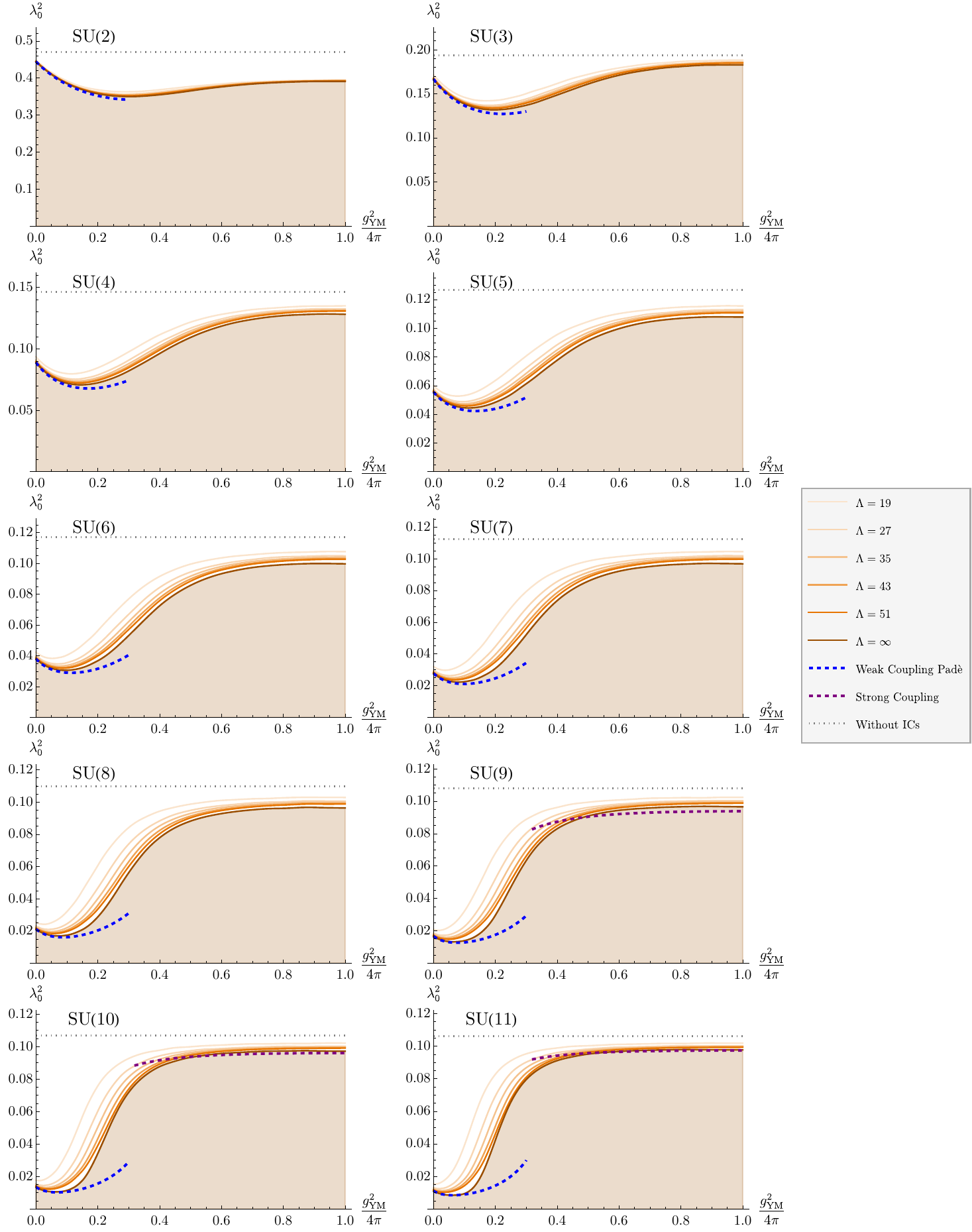}
	\caption{Bootstrap bounds on the OPE coefficient of the lowest unprotected scalar operator for $2\leq N\leq 11$, obtained as in Figure~\ref{fig:ope}. For $N\geq 9$, we compare with the large $N$ expansion as in Figure~\ref{fig:ope}. For $N<9$, this expansion is not sufficiently converged to be of use.}
	\label{fig:ope_all}
\end{figure}

Generically, in the linear programming approach to the bootstrap, the functionals obtained can be negative at scaling dimensions where positivity is not imposed. In practice, our functionals will become slightly negative near where a true positive functional would have a zero. As we make the grid finer, the extremal functional approaches a true positive functional (up to the maximum twist; see Section \ref{setup2} for a discussion of asymptotic positivity). We find that the bounds we obtain change very little as the grid is made finer, so that the coarseness is not a significant source of error in our results.

For all of our results, we obtain bootstrap bounds at $\Lambda = 19, 27, 35, 43,$ and $51$, using binary search with a tolerance of $10^{-4}$. We then extrapolate these results to $\Lambda\to\infty$ using a polynomial in $\frac{1}{\Lambda}$. Figure~\ref{fig:extrapolation} gives an example of this extrapolation for the upper bound on the scaling dimension in the $\grSU(10)$ theory at $\frac{g_\text{YM}^2}{4\pi} = \frac{1}{9}$.

In Section~\ref{results}, we show selected results for certain values of $N$. In Figures~\ref{fig:dimension_all} and \ref{fig:ope_all}, we give our results for all $2\leq N\leq 11$.

\bibliographystyle{ssg}
\bibliography{4dIntBoot_updated}

\begingroup\raggedright\begin{thebibliography}{100}

\bibitem{Grisaru:1980nk}
M.~T. Grisaru, M.~Rocek, and W.~Siegel, ``{Zero Three Loop beta Function in N=4
  Superyang-Mills Theory},'' {\em Phys. Rev. Lett.} {\bf 45} (1980) 1063--1066.

\bibitem{Grisaru:1980jc}
M.~T. Grisaru, M.~Rocek, and W.~Siegel, ``{Superloops 3, Beta 0: A Calculation
  in $N=4$ {Yang-Mills} Theory},'' {\em Nucl. Phys. B} {\bf 183} (1981)
  141--156.

\bibitem{Caswell:1980ru}
W.~E. Caswell and D.~Zanon, ``{Zero Three Loop Beta Function in the $N=4$
  Supersymmetric {Yang-Mills} Theory},'' {\em Nucl. Phys. B} {\bf 182} (1981)
  125.

\bibitem{Caswell:1980yi}
W.~E. Caswell and D.~Zanon, ``{Vanishing Three Loop Beta Function in $N=4$
  Supersymmetric {Yang-Mills} Theory},'' {\em Phys. Lett. B} {\bf 100} (1981)
  152--156.

\bibitem{Sohnius:1981sn}
M.~F. Sohnius and P.~C. West, ``{Conformal Invariance in N=4 Supersymmetric
  Yang-Mills Theory},'' {\em Phys. Lett. B} {\bf 100} (1981) 245.

\bibitem{Minwalla:1997ka}
S.~Minwalla, ``{Restrictions imposed by superconformal invariance on quantum
  field theories},'' {\em Adv. Theor. Math. Phys.} {\bf 2} (1998) 781--846,
  \href{https://arxiv.org/abs/hep-th/9712074}{{\tt hep-th/9712074}}.

\bibitem{Dolan:2002zh}
F.~A. Dolan and H.~Osborn, ``{On short and semi-short representations for
  four-dimensional superconformal symmetry},'' {\em Annals Phys.} {\bf 307}
  (2003) 41--89, \href{https://arxiv.org/abs/hep-th/0209056}{{\tt
  hep-th/0209056}}.

\bibitem{Cordova:2016emh}
C.~Cordova, T.~T. Dumitrescu, and K.~Intriligator, ``{Multiplets of
  Superconformal Symmetry in Diverse Dimensions},'' {\em JHEP} {\bf 03} (2019)
  163, \href{https://arxiv.org/abs/1612.00809}{{\tt 1612.00809}}.

\bibitem{Lee:1998bxa}
S.~Lee, S.~Minwalla, M.~Rangamani, and N.~Seiberg, ``{Three point functions of
  chiral operators in D = 4, N=4 SYM at large N},'' {\em Adv. Theor. Math.
  Phys.} {\bf 2} (1998) 697--718,
  \href{https://arxiv.org/abs/hep-th/9806074}{{\tt hep-th/9806074}}.

\bibitem{Konishi:1983hf}
K.~Konishi, ``{Anomalous Supersymmetry Transformation of Some Composite
  Operators in SQCD},'' {\em Phys. Lett. B} {\bf 135} (1984) 439--444.

\bibitem{Velizhanin:2009gv}
V.~N. Velizhanin, ``{The Non-planar contribution to the four-loop universal
  anomalous dimension in N=4 Supersymmetric Yang-Mills theory},'' {\em JETP
  Lett.} {\bf 89} (2009) 593--596, \href{https://arxiv.org/abs/0902.4646}{{\tt
  0902.4646}}.

\bibitem{Eden:2012rr}
B.~Eden, ``{Three-loop universal structure constants in N=4 susy Yang-Mills
  theory},'' \href{https://arxiv.org/abs/1207.3112}{{\tt 1207.3112}}.

\bibitem{Fleury:2019ydf}
T.~Fleury and R.~Pereira, ``{Non-planar data of $ \mathcal{N} $ = 4 SYM},''
  {\em JHEP} {\bf 03} (2020) 003, \href{https://arxiv.org/abs/1910.09428}{{\tt
  1910.09428}}.

\bibitem{Eden:2016aqo}
B.~Eden and F.~Paul, ``{Half-BPS half-BPS twist two at four loops in N=4
  SYM},'' \href{https://arxiv.org/abs/1608.04222}{{\tt 1608.04222}}.

\bibitem{Goncalves:2016vir}
V.~Gon\c{c}alves, ``{Extracting OPE coefficient of Konishi at four loops},''
  {\em JHEP} {\bf 03} (2017) 079, \href{https://arxiv.org/abs/1607.02195}{{\tt
  1607.02195}}.

\bibitem{Maldacena:1997re}
J.~M. Maldacena, ``{The Large $N$ limit of superconformal field theories and
  supergravity},'' {\em Int. J. Theor. Phys.} {\bf 38} (1999) 1113--1133,
  \href{https://arxiv.org/abs/hep-th/9711200}{{\tt hep-th/9711200}}. [Adv.
  Theor. Math. Phys.2,231(1998)].

\bibitem{Gubser:1998bc}
S.~S. Gubser, I.~R. Klebanov, and A.~M. Polyakov, ``{Gauge theory correlators
  from noncritical string theory},'' {\em Phys. Lett.} {\bf B428} (1998)
  105--114, \href{https://arxiv.org/abs/hep-th/9802109}{{\tt hep-th/9802109}}.

\bibitem{Witten:1998qj}
E.~Witten, ``{Anti-de Sitter space and holography},'' {\em Adv. Theor. Math.
  Phys.} {\bf 2} (1998) 253--291,
  \href{https://arxiv.org/abs/hep-th/9802150}{{\tt hep-th/9802150}}.

\bibitem{Aharony:1999ti}
O.~Aharony, S.~S. Gubser, J.~M. Maldacena, H.~Ooguri, and Y.~Oz, ``{Large N
  field theories, string theory and gravity},'' {\em Phys. Rept.} {\bf 323}
  (2000) 183--386, \href{https://arxiv.org/abs/hep-th/9905111}{{\tt
  hep-th/9905111}}.

\bibitem{DHoker:2002nbb}
E.~D'Hoker and D.~Z. Freedman, ``{Supersymmetric gauge theories and the AdS /
  CFT correspondence},'' in {\em {Theoretical Advanced Study Institute in
  Elementary Particle Physics (TASI 2001): Strings, Branes and EXTRA
  Dimensions}}, pp.~3--158, 1, 2002.
\newblock \href{https://arxiv.org/abs/hep-th/0201253}{{\tt hep-th/0201253}}.

\bibitem{Klebanov:2000me}
I.~R. Klebanov, ``{TASI lectures: Introduction to the AdS / CFT
  correspondence},'' in {\em {Theoretical Advanced Study Institute in
  Elementary Particle Physics (TASI 99): Strings, Branes, and Gravity}},
  pp.~615--650, 9, 2000.
\newblock \href{https://arxiv.org/abs/hep-th/0009139}{{\tt hep-th/0009139}}.

\bibitem{Polchinski:2010hw}
J.~Polchinski, ``{Introduction to Gauge/Gravity Duality},'' in {\em
  {Theoretical Advanced Study Institute in Elementary Particle Physics}:
  {String theory and its Applications: From meV to the Planck Scale}},
  pp.~3--46, 10, 2010.
\newblock \href{https://arxiv.org/abs/1010.6134}{{\tt 1010.6134}}.

\bibitem{Kim:1985ez}
H.~J. Kim, L.~J. Romans, and P.~van Nieuwenhuizen, ``{The Mass Spectrum of
  Chiral N=2 D=10 Supergravity on S**5},'' {\em Phys. Rev. D} {\bf 32} (1985)
  389.

\bibitem{Gubser:2002tv}
S.~S. Gubser, I.~R. Klebanov, and A.~M. Polyakov, ``{A Semiclassical limit of
  the gauge / string correspondence},'' {\em Nucl. Phys. B} {\bf 636} (2002)
  99--114, \href{https://arxiv.org/abs/hep-th/0204051}{{\tt hep-th/0204051}}.

\bibitem{tHooft:1973alw}
G.~'t~Hooft, ``{A Planar Diagram Theory for Strong Interactions},'' {\em Nucl.
  Phys. B} {\bf 72} (1974) 461.

\bibitem{Minahan:2002ve}
J.~A. Minahan and K.~Zarembo, ``{The Bethe ansatz for N=4 superYang-Mills},''
  {\em JHEP} {\bf 03} (2003) 013,
  \href{https://arxiv.org/abs/hep-th/0212208}{{\tt hep-th/0212208}}.

\bibitem{Bena:2003wd}
I.~Bena, J.~Polchinski, and R.~Roiban, ``{Hidden symmetries of the AdS(5) x
  S**5 superstring},'' {\em Phys. Rev. D} {\bf 69} (2004) 046002,
  \href{https://arxiv.org/abs/hep-th/0305116}{{\tt hep-th/0305116}}.

\bibitem{Beisert:2003tq}
N.~Beisert, C.~Kristjansen, and M.~Staudacher, ``{The Dilatation operator of
  conformal N=4 superYang-Mills theory},'' {\em Nucl. Phys. B} {\bf 664} (2003)
  131--184, \href{https://arxiv.org/abs/hep-th/0303060}{{\tt hep-th/0303060}}.

\bibitem{Kazakov:2004qf}
V.~A. Kazakov, A.~Marshakov, J.~A. Minahan, and K.~Zarembo,
  ``{Classical/quantum integrability in AdS/CFT},'' {\em JHEP} {\bf 05} (2004)
  024, \href{https://arxiv.org/abs/hep-th/0402207}{{\tt hep-th/0402207}}.

\bibitem{Beisert:2005bm}
N.~Beisert, V.~A. Kazakov, K.~Sakai, and K.~Zarembo, ``{The Algebraic curve of
  classical superstrings on AdS(5) x S**5},'' {\em Commun. Math. Phys.} {\bf
  263} (2006) 659--710, \href{https://arxiv.org/abs/hep-th/0502226}{{\tt
  hep-th/0502226}}.

\bibitem{Staudacher:2004tk}
M.~Staudacher, ``{The Factorized S-matrix of CFT/AdS},'' {\em JHEP} {\bf 05}
  (2005) 054, \href{https://arxiv.org/abs/hep-th/0412188}{{\tt
  hep-th/0412188}}.

\bibitem{Beisert:2005tm}
N.~Beisert, ``{The SU(2|2) dynamic S-matrix},'' {\em Adv. Theor. Math. Phys.}
  {\bf 12} (2008) 945--979, \href{https://arxiv.org/abs/hep-th/0511082}{{\tt
  hep-th/0511082}}.

\bibitem{Arutyunov:2004vx}
G.~Arutyunov, S.~Frolov, and M.~Staudacher, ``{Bethe ansatz for quantum
  strings},'' {\em JHEP} {\bf 10} (2004) 016,
  \href{https://arxiv.org/abs/hep-th/0406256}{{\tt hep-th/0406256}}.

\bibitem{Beisert:2006ib}
N.~Beisert, R.~Hernandez, and E.~Lopez, ``{A Crossing-symmetric phase for
  AdS(5) x S**5 strings},'' {\em JHEP} {\bf 11} (2006) 070,
  \href{https://arxiv.org/abs/hep-th/0609044}{{\tt hep-th/0609044}}.

\bibitem{Beisert:2010jr}
N.~Beisert {\em et.~al.}, ``{Review of AdS/CFT Integrability: An Overview},''
  {\em Lett. Math. Phys.} {\bf 99} (2012) 3--32,
  \href{https://arxiv.org/abs/1012.3982}{{\tt 1012.3982}}.

\bibitem{Gromov:2023hzc}
N.~Gromov, A.~Hegedus, J.~Julius, and N.~Sokolova, ``{Fast QSC Solver: tool for
  systematic study of $N=4$ Super-Yang-Mills spectrum},''
  \href{https://arxiv.org/abs/2306.12379}{{\tt 2306.12379}}.

\bibitem{Harmark:2017yrv}
T.~Harmark and M.~Wilhelm, ``{Hagedorn Temperature of AdS$_5$/CFT$_4$ via
  Integrability},'' {\em Phys. Rev. Lett.} {\bf 120} (2018), no.~7 071605,
  \href{https://arxiv.org/abs/1706.03074}{{\tt 1706.03074}}.

\bibitem{Urbach:2022xzw}
E.~Y. Urbach, ``{String stars in anti de Sitter space},'' {\em JHEP} {\bf 04}
  (2022) 072, \href{https://arxiv.org/abs/2202.06966}{{\tt 2202.06966}}.

\bibitem{Harmark:2021qma}
T.~Harmark and M.~Wilhelm, ``{Solving the Hagedorn temperature of
  AdS$_{5}$/CFT$_{4}$ via the Quantum Spectral Curve: chemical potentials and
  deformations},'' {\em JHEP} {\bf 07} (2022) 136,
  \href{https://arxiv.org/abs/2109.09761}{{\tt 2109.09761}}.

\bibitem{Bigazzi:2023oqm}
F.~Bigazzi, T.~Canneti, and W.~M\"uck, ``{Semiclassical quantization of the
  superstring and Hagedorn temperature},'' {\em JHEP} {\bf 08} (2023) 185,
  \href{https://arxiv.org/abs/2306.00588}{{\tt 2306.00588}}.

\bibitem{Ekhammar:2023glu}
S.~Ekhammar, J.~A. Minahan, and C.~Thull, ``{The asymptotic form of the
  Hagedorn temperature in planar super Yang-Mills},'' {\em J. Phys. A} {\bf 56}
  (2023), no.~43 435401, \href{https://arxiv.org/abs/2306.09883}{{\tt
  2306.09883}}.

\bibitem{Bigazzi:2023hxt}
F.~Bigazzi, T.~Canneti, and A.~L. Cotrone, ``{Higher order corrections to the
  Hagedorn temperature at strong coupling},'' {\em JHEP} {\bf 10} (2023) 056,
  \href{https://arxiv.org/abs/2306.17126}{{\tt 2306.17126}}.

\bibitem{Alday:2023mvu}
L.~F. Alday and T.~Hansen, ``{The AdS Virasoro-Shapiro amplitude},'' {\em JHEP}
  {\bf 10} (2023) 023, \href{https://arxiv.org/abs/2306.12786}{{\tt
  2306.12786}}.

\bibitem{Alday:2023jdk}
L.~F. Alday, T.~Hansen, and J.~A. Silva, ``{Emergent Worldsheet for the AdS
  Virasoro-Shapiro Amplitude},'' {\em Phys. Rev. Lett.} {\bf 131} (2023),
  no.~16 161603, \href{https://arxiv.org/abs/2305.03593}{{\tt 2305.03593}}.

\bibitem{Alday:2023flc}
L.~F. Alday, T.~Hansen, and J.~A. Silva, ``{On the spectrum and structure
  constants of short operators in N=4 SYM at strong coupling},'' {\em JHEP}
  {\bf 08} (2023) 214, \href{https://arxiv.org/abs/2303.08834}{{\tt
  2303.08834}}.

\bibitem{Alday:2022xwz}
L.~F. Alday, T.~Hansen, and J.~A. Silva, ``{AdS Virasoro-Shapiro from
  single-valued periods},'' {\em JHEP} {\bf 12} (2022) 010,
  \href{https://arxiv.org/abs/2209.06223}{{\tt 2209.06223}}.

\bibitem{Alday:2022uxp}
L.~F. Alday, T.~Hansen, and J.~A. Silva, ``{AdS Virasoro-Shapiro from
  dispersive sum rules},'' {\em JHEP} {\bf 10} (2022) 036,
  \href{https://arxiv.org/abs/2204.07542}{{\tt 2204.07542}}.

\bibitem{Bargheer:2017nne}
T.~Bargheer, J.~Caetano, T.~Fleury, S.~Komatsu, and P.~Vieira, ``{Handling
  Handles: Nonplanar Integrability in $\mathcal{N}=4$ Supersymmetric Yang-Mills
  Theory},'' {\em Phys. Rev. Lett.} {\bf 121} (2018), no.~23 231602,
  \href{https://arxiv.org/abs/1711.05326}{{\tt 1711.05326}}.

\bibitem{Basso:2015zoa}
B.~Basso, S.~Komatsu, and P.~Vieira, ``{Structure Constants and Integrable
  Bootstrap in Planar N=4 SYM Theory},''
  \href{https://arxiv.org/abs/1505.06745}{{\tt 1505.06745}}.

\bibitem{Cavaglia:2022qpg}
A.~Cavagli\`a, N.~Gromov, J.~Julius, and M.~Preti, ``{Bootstrability in defect
  CFT: integrated correlators and sharper bounds},'' {\em JHEP} {\bf 05} (2022)
  164, \href{https://arxiv.org/abs/2203.09556}{{\tt 2203.09556}}.

\bibitem{Cavaglia:2021bnz}
A.~Cavagli\`a, N.~Gromov, J.~Julius, and M.~Preti, ``{Integrability and
  conformal bootstrap: One dimensional defect conformal field theory},'' {\em
  Phys. Rev. D} {\bf 105} (2022), no.~2 L021902,
  \href{https://arxiv.org/abs/2107.08510}{{\tt 2107.08510}}.

\bibitem{Cavaglia:2022yvv}
A.~Cavagli\`a, N.~Gromov, J.~Julius, and M.~Preti, ``{Integrated correlators
  from integrability: Maldacena-Wilson line in $ \mathcal{N} $ = 4 SYM},'' {\em
  JHEP} {\bf 04} (2023) 026, \href{https://arxiv.org/abs/2211.03203}{{\tt
  2211.03203}}.

\bibitem{Caron-Huot:2022sdy}
S.~Caron-Huot, F.~Coronado, A.-K. Trinh, and Z.~Zahraee, ``{Bootstrapping $
  \mathcal{N} $ = 4 sYM correlators using integrability},'' {\em JHEP} {\bf 02}
  (2023) 083, \href{https://arxiv.org/abs/2207.01615}{{\tt 2207.01615}}.

\bibitem{Korchemsky:2015cyx}
G.~P. Korchemsky, ``{On level crossing in conformal field theories},'' {\em
  JHEP} {\bf 03} (2016) 212, \href{https://arxiv.org/abs/1512.05362}{{\tt
  1512.05362}}.

\bibitem{Chester:2019wfx}
S.~M. Chester, ``{Weizmann Lectures on the Numerical Conformal Bootstrap},''
  \href{https://arxiv.org/abs/1907.05147}{{\tt 1907.05147}}.

\bibitem{Poland:2018epd}
D.~Poland, S.~Rychkov, and A.~Vichi, ``{The Conformal Bootstrap: Theory,
  Numerical Techniques, and Applications},'' {\em Rev. Mod. Phys.} {\bf 91}
  (2019) 015002, \href{https://arxiv.org/abs/1805.04405}{{\tt 1805.04405}}.

\bibitem{Poland:2022qrs}
D.~Poland and D.~Simmons-Duffin, ``{Snowmass White Paper: The Numerical
  Conformal Bootstrap},'' in {\em {Snowmass 2021}}, 3, 2022.
\newblock \href{https://arxiv.org/abs/2203.08117}{{\tt 2203.08117}}.

\bibitem{Rychkov:2023wsd}
S.~Rychkov and N.~Su, ``{New Developments in the Numerical Conformal
  Bootstrap},'' \href{https://arxiv.org/abs/2311.15844}{{\tt 2311.15844}}.

\bibitem{Rattazzi:2008pe}
R.~Rattazzi, V.~S. Rychkov, E.~Tonni, and A.~Vichi, ``{Bounding scalar operator
  dimensions in 4D CFT},'' {\em JHEP} {\bf 0812} (2008) 031,
  \href{https://arxiv.org/abs/0807.0004}{{\tt 0807.0004}}.

\bibitem{Binder:2019jwn}
D.~J. Binder, S.~M. Chester, S.~S. Pufu, and Y.~Wang, ``{$\mathcal{N}=4$
  Super-Yang-Mills Correlators at Strong Coupling from String Theory and
  Localization},'' {\em JHEP} {\bf 2019} (2019), no.~12
  \href{https://arxiv.org/abs/1902.06263}{{\tt 1902.06263}}.

\bibitem{Chester:2019jas}
S.~M. Chester, M.~B. Green, S.~S. Pufu, Y.~Wang, and C.~Wen, ``{Modular
  invariance in superstring theory from $ \mathcal{N} $ = 4
  super-Yang-Mills},'' {\em JHEP} {\bf 11} (2020) 016,
  \href{https://arxiv.org/abs/1912.13365}{{\tt 1912.13365}}.

\bibitem{Chester:2020vyz}
S.~M. Chester, M.~B. Green, S.~S. Pufu, Y.~Wang, and C.~Wen, ``{New modular
  invariants in $ \mathcal{N} $ = 4 Super-Yang-Mills theory},'' {\em JHEP} {\bf
  04} (2021) 212, \href{https://arxiv.org/abs/2008.02713}{{\tt 2008.02713}}.

\bibitem{Chester:2020dja}
S.~M. Chester and S.~S. Pufu, ``{Far beyond the planar limit in
  strongly-coupled $ \mathcal{N} $ = 4 SYM},'' {\em JHEP} {\bf 01} (2021) 103,
  \href{https://arxiv.org/abs/2003.08412}{{\tt 2003.08412}}.

\bibitem{Chester:2019pvm}
S.~M. Chester, ``{Genus-2 holographic correlator on AdS$_{5}$\texttimes{}
  S$^{5}$ from localization},'' {\em JHEP} {\bf 04} (2020) 193,
  \href{https://arxiv.org/abs/1908.05247}{{\tt 1908.05247}}.

\bibitem{Heemskerk:2009pn}
I.~Heemskerk, J.~Penedones, J.~Polchinski, and J.~Sully, ``{Holography from
  Conformal Field Theory},'' {\em JHEP} {\bf 10} (2009) 079,
  \href{https://arxiv.org/abs/0907.0151}{{\tt 0907.0151}}.

\bibitem{Rastelli:2017udc}
L.~Rastelli and X.~Zhou, ``{How to Succeed at Holographic Correlators Without
  Really Trying},'' {\em JHEP} {\bf 2018} (2017), no.~4
  \href{https://arxiv.org/abs/1710.05923}{{\tt 1710.05923}}.

\bibitem{Pestun:2007rz}
V.~Pestun, ``{Localization of gauge theory on a four-sphere and supersymmetric
  Wilson loops},'' {\em Commun. Math. Phys.} {\bf 313} (2012) 71--129,
  \href{https://arxiv.org/abs/0712.2824}{{\tt 0712.2824}}.

\bibitem{Beem:2013qxa}
C.~Beem, L.~Rastelli, and B.~C. van Rees, ``{The $\mathcal N=4$ Superconformal
  Bootstrap},'' {\em Phys.Rev.Lett.} {\bf 111} (2013), no.~7 071601,
  \href{https://arxiv.org/abs/1304.1803}{{\tt 1304.1803}}.

\bibitem{Alday:2022ldo}
L.~F. Alday and S.~M. Chester, ``{Pure Anti\textendash{}de Sitter Supergravity
  and the Conformal Bootstrap},'' {\em Phys. Rev. Lett.} {\bf 129} (2022),
  no.~21 211601, \href{https://arxiv.org/abs/2207.05085}{{\tt 2207.05085}}.

\bibitem{ElShowk:2012ht}
S.~El-Showk, M.~F. Paulos, D.~Poland, S.~Rychkov, D.~Simmons-Duffin, {\em
  et.~al.}, ``{Solving the 3D Ising Model with the Conformal Bootstrap},'' {\em
  Phys.Rev.} {\bf D86} (2012) 025022,
  \href{https://arxiv.org/abs/1203.6064}{{\tt 1203.6064}}.

\bibitem{Kos:2013tga}
F.~Kos, D.~Poland, and D.~Simmons-Duffin, ``{Bootstrapping the $O(N)$ vector
  models},'' {\em JHEP} {\bf 06} (2014) 091,
  \href{https://arxiv.org/abs/1307.6856}{{\tt 1307.6856}}.

\bibitem{Kos:2015mba}
F.~Kos, D.~Poland, D.~Simmons-Duffin, and A.~Vichi, ``{Bootstrapping the O(N)
  Archipelago},'' {\em JHEP} {\bf 11} (2015) 106,
  \href{https://arxiv.org/abs/1504.07997}{{\tt 1504.07997}}.

\bibitem{Kos:2016ysd}
F.~Kos, D.~Poland, D.~Simmons-Duffin, and A.~Vichi, ``{Precision Islands in the
  Ising and $O(N)$ Models},'' {\em JHEP} {\bf 08} (2016) 036,
  \href{https://arxiv.org/abs/1603.04436}{{\tt 1603.04436}}.

\bibitem{Chester:2020iyt}
S.~M. Chester, W.~Landry, J.~Liu, D.~Poland, D.~Simmons-Duffin, N.~Su, and
  A.~Vichi, ``{Bootstrapping Heisenberg Magnets and their Cubic Instability},''
  \href{https://arxiv.org/abs/2011.14647}{{\tt 2011.14647}}.

\bibitem{Chester:2019ifh}
S.~M. Chester, W.~Landry, J.~Liu, D.~Poland, D.~Simmons-Duffin, N.~Su, and
  A.~Vichi, ``{Carving out OPE space and precise $O(2)$ model critical
  exponents},'' {\em JHEP} {\bf 06} (2020) 142,
  \href{https://arxiv.org/abs/1912.03324}{{\tt 1912.03324}}.

\bibitem{Chester:2016wrc}
S.~M. Chester and S.~S. Pufu, ``{Towards bootstrapping QED$_{3}$},'' {\em JHEP}
  {\bf 08} (2016) 019, \href{https://arxiv.org/abs/1601.03476}{{\tt
  1601.03476}}.

\bibitem{Albayrak:2021xtd}
S.~Albayrak, R.~S. Erramilli, Z.~Li, D.~Poland, and Y.~Xin, ``{Bootstrapping
  $N_f$=4 conformal QED$_3$},'' {\em Phys. Rev. D} {\bf 105} (2022), no.~8
  085008, \href{https://arxiv.org/abs/2112.02106}{{\tt 2112.02106}}.

\bibitem{Chester:2023njo}
S.~M. Chester and N.~Su, ``{Bootstrapping Deconfined Quantum Tricriticality},''
  \href{https://arxiv.org/abs/2310.08343}{{\tt 2310.08343}}.

\bibitem{Chester:2022hzt}
S.~M. Chester and N.~Su, ``{Upper critical dimension of the 3-state Potts
  model},'' \href{https://arxiv.org/abs/2210.09091}{{\tt 2210.09091}}.

\bibitem{Beem:2016wfs}
C.~Beem, L.~Rastelli, and B.~C. van Rees, ``{More ${\mathcal N}=4$
  superconformal bootstrap},'' {\em Phys. Rev.} {\bf D96} (2017), no.~4 046014,
  \href{https://arxiv.org/abs/1612.02363}{{\tt 1612.02363}}.

\bibitem{Belitsky:2014zha}
A.~V. Belitsky, S.~Hohenegger, G.~P. Korchemsky, and E.~Sokatchev, ``{N=4
  superconformal Ward identities for correlation functions},'' {\em Nucl. Phys.
  B} {\bf 904} (2016) 176--215, \href{https://arxiv.org/abs/1409.2502}{{\tt
  1409.2502}}.

\bibitem{Beem:2013sza}
C.~Beem, M.~Lemos, P.~Liendo, W.~Peelaers, L.~Rastelli, and B.~C. van Rees,
  ``{Infinite Chiral Symmetry in Four Dimensions},'' {\em Commun. Math. Phys.}
  {\bf 336} (2015), no.~3 1359--1433,
  \href{https://arxiv.org/abs/1312.5344}{{\tt 1312.5344}}.

\bibitem{Chester:2021aun}
S.~M. Chester, R.~Dempsey, and S.~S. Pufu, ``{Bootstrapping $ \mathcal{N} $ = 4
  super-Yang-Mills on the conformal manifold},'' {\em JHEP} {\bf 01} (2023)
  038, \href{https://arxiv.org/abs/2111.07989}{{\tt 2111.07989}}.

\bibitem{Alday:2023pet}
L.~F. Alday, S.~M. Chester, D.~Dorigoni, M.~B. Green, and C.~Wen, ``{Relations
  between integrated correlators in $\mathcal{N}=4$ Supersymmetric Yang--Mills
  Theory},'' \href{https://arxiv.org/abs/2310.12322}{{\tt 2310.12322}}.

\bibitem{Nekrasov:2002qd}
N.~A. Nekrasov, ``{Seiberg-Witten prepotential from instanton counting},'' {\em
  Adv. Theor. Math. Phys.} {\bf 7} (2003), no.~5 831--864,
  \href{https://arxiv.org/abs/hep-th/0206161}{{\tt hep-th/0206161}}.

\bibitem{Nekrasov:2003rj}
N.~Nekrasov and A.~Okounkov, ``{Seiberg-Witten theory and random partitions},''
  {\em Prog. Math.} {\bf 244} (2006) 525--596,
  \href{https://arxiv.org/abs/hep-th/0306238}{{\tt hep-th/0306238}}.

\bibitem{Green:2014yxa}
M.~B. Green, S.~D. Miller, and P.~Vanhove, ``{$SL(2, \mathbb{Z})$-invariance
  and D-instanton contributions to the $D^6 R^4$ interaction},'' {\em Commun.
  Num. Theor. Phys.} {\bf 09} (2015) 307--344,
  \href{https://arxiv.org/abs/1404.2192}{{\tt 1404.2192}}.

\bibitem{Gromov:2013pga}
N.~Gromov, V.~Kazakov, S.~Leurent, and D.~Volin, ``{Quantum Spectral Curve for
  Planar $\mathcal{N} = 4$ Super-Yang-Mills Theory},'' {\em Phys. Rev. Lett.}
  {\bf 112} (2014), no.~1 011602, \href{https://arxiv.org/abs/1305.1939}{{\tt
  1305.1939}}.

\bibitem{Gromov:2014caa}
N.~Gromov, V.~Kazakov, S.~Leurent, and D.~Volin, ``{Quantum spectral curve for
  arbitrary state/operator in AdS$_{5}$/CFT$_{4}$},'' {\em JHEP} {\bf 09}
  (2015) 187, \href{https://arxiv.org/abs/1405.4857}{{\tt 1405.4857}}.

\bibitem{Gromov:2011bz}
N.~Gromov and S.~Valatka, ``{Deeper Look into Short Strings},'' {\em JHEP} {\bf
  03} (2012) 058, \href{https://arxiv.org/abs/1109.6305}{{\tt 1109.6305}}.

\bibitem{Gromov:2014bva}
N.~Gromov, F.~Levkovich-Maslyuk, G.~Sizov, and S.~Valatka, ``{Quantum spectral
  curve at work: from small spin to strong coupling in $ \mathcal{N} $ = 4
  SYM},'' {\em JHEP} {\bf 07} (2014) 156,
  \href{https://arxiv.org/abs/1402.0871}{{\tt 1402.0871}}.

\bibitem{Basso:2011rs}
B.~Basso, ``{An exact slope for AdS/CFT},''
  \href{https://arxiv.org/abs/1109.3154}{{\tt 1109.3154}}.

\bibitem{Gromov:2011de}
N.~Gromov, D.~Serban, I.~Shenderovich, and D.~Volin, ``{Quantum folded string
  and integrability: From finite size effects to Konishi dimension},'' {\em
  JHEP} {\bf 08} (2011) 046, \href{https://arxiv.org/abs/1102.1040}{{\tt
  1102.1040}}.

\bibitem{Aprile:2017bgs}
F.~Aprile, J.~M. Drummond, P.~Heslop, and H.~Paul, ``{Quantum Gravity from
  Conformal Field Theory},'' {\em JHEP} {\bf 01} (2018) 035,
  \href{https://arxiv.org/abs/1706.02822}{{\tt 1706.02822}}.

\bibitem{Alday:2017xua}
L.~F. Alday and A.~Bissi, ``{Loop Corrections to Supergravity on $AdS_5 \times
  S^5$},'' {\em Phys. Rev. Lett.} {\bf 119} (2017), no.~17 171601,
  \href{https://arxiv.org/abs/1706.02388}{{\tt 1706.02388}}.

\bibitem{Alday:2021peq}
L.~F. Alday, S.~M. Chester, and T.~Hansen, ``{Modular invariant holographic
  correlators for $\mathcal{N}=4$ SYM with general gauge group},''
  \href{https://arxiv.org/abs/2110.13106}{{\tt 2110.13106}}.

\bibitem{Hogervorst:2013sma}
M.~Hogervorst and S.~Rychkov, ``{Radial Coordinates for Conformal Blocks},''
  {\em Phys.Rev.} {\bf D87} (2013), no.~10 106004,
  \href{https://arxiv.org/abs/1303.1111}{{\tt 1303.1111}}.

\bibitem{Lin:2015wcg}
Y.-H. Lin, S.-H. Shao, D.~Simmons-Duffin, Y.~Wang, and X.~Yin, ``{$ \mathcal{N}
  $ = 4 superconformal bootstrap of the K3 CFT},'' {\em JHEP} {\bf 05} (2017)
  126, \href{https://arxiv.org/abs/1511.04065}{{\tt 1511.04065}}.

\bibitem{Brown:2023cpz}
A.~Brown, C.~Wen, and H.~Xie, ``{Laplace-difference equation for integrated
  correlators of operators with general charges in $ \mathcal{N} $ = 4 SYM},''
  {\em JHEP} {\bf 06} (2023) 066, \href{https://arxiv.org/abs/2303.13195}{{\tt
  2303.13195}}.

\bibitem{Brown:2023why}
A.~Brown, C.~Wen, and H.~Xie, ``{Generating functions and large-charge
  expansion of integrated correlators in N = 4 supersymmetric Yang-Mills
  theory},'' {\em JHEP} {\bf 07} (2023) 129,
  \href{https://arxiv.org/abs/2303.17570}{{\tt 2303.17570}}.

\bibitem{Paul:2023rka}
H.~Paul, E.~Perlmutter, and H.~Raj, ``{Exact large charge in $ \mathcal{N} $ =
  4 SYM and semiclassical string theory},'' {\em JHEP} {\bf 08} (2023) 078,
  \href{https://arxiv.org/abs/2303.13207}{{\tt 2303.13207}}.

\bibitem{Paul:2022piq}
H.~Paul, E.~Perlmutter, and H.~Raj, ``{Integrated correlators in $ \mathcal{N}
  $ = 4 SYM via SL(2, Z) spectral theory},'' {\em JHEP} {\bf 01} (2023) 149,
  \href{https://arxiv.org/abs/2209.06639}{{\tt 2209.06639}}.

\bibitem{Kos:2014bka}
F.~Kos, D.~Poland, and D.~Simmons-Duffin, ``{Bootstrapping Mixed Correlators in
  the 3D Ising Model},'' {\em JHEP} {\bf 11} (2014) 109,
  \href{https://arxiv.org/abs/1406.4858}{{\tt 1406.4858}}.

\bibitem{Alday:2021ymb}
L.~F. Alday, S.~M. Chester, and H.~Raj, ``{ABJM at Strong Coupling from
  M-theory, Localization, and Lorentzian Inversion},''
  \href{https://arxiv.org/abs/2107.10274}{{\tt 2107.10274}}.

\bibitem{Bargheer:2018jvq}
T.~Bargheer, J.~Caetano, T.~Fleury, S.~Komatsu, and P.~Vieira, ``{Handling
  handles. Part II. Stratification and data analysis},'' {\em JHEP} {\bf 11}
  (2018) 095, \href{https://arxiv.org/abs/1809.09145}{{\tt 1809.09145}}.

\bibitem{Chester:2022sqb}
S.~M. Chester, ``{Bootstrapping 4d $ \mathcal{N} $ = 2 gauge theories: the case
  of SQCD},'' {\em JHEP} {\bf 01} (2023) 107,
  \href{https://arxiv.org/abs/2205.12978}{{\tt 2205.12978}}.

\bibitem{Behan:2023fqq}
C.~Behan, S.~M. Chester, and P.~Ferrero, ``{Gluon scattering in AdS at finite
  string coupling from localization},''
  \href{https://arxiv.org/abs/2305.01016}{{\tt 2305.01016}}.

\bibitem{Chester:2014fya}
S.~M. Chester, J.~Lee, S.~S. Pufu, and R.~Yacoby, ``{The $ \mathcal{N}=8 $
  superconformal bootstrap in three dimensions},'' {\em JHEP} {\bf 09} (2014)
  143, \href{https://arxiv.org/abs/1406.4814}{{\tt 1406.4814}}.

\bibitem{Chester:2014mea}
S.~M. Chester, J.~Lee, S.~S. Pufu, and R.~Yacoby, ``{Exact Correlators of BPS
  Operators from the 3d Superconformal Bootstrap},'' {\em JHEP} {\bf 03} (2015)
  130, \href{https://arxiv.org/abs/1412.0334}{{\tt 1412.0334}}.

\bibitem{Agmon:2017xes}
N.~B. Agmon, S.~M. Chester, and S.~S. Pufu, ``{Solving M-theory with the
  Conformal Bootstrap},'' {\em JHEP} {\bf 2018} (2017), no.~6
  \href{https://arxiv.org/abs/1711.07343}{{\tt 1711.07343}}.

\bibitem{Agmon:2019imm}
N.~B. Agmon, S.~M. Chester, and S.~S. Pufu, ``{The M-theory Archipelago},''
  {\em JHEP} {\bf 02} (2020) 010, \href{https://arxiv.org/abs/1907.13222}{{\tt
  1907.13222}}.

\bibitem{Binder:2020ckj}
D.~J. Binder, S.~M. Chester, M.~Jerdee, and S.~S. Pufu, ``{The 3d
  $\mathcal{N}=6$ Bootstrap: From Higher Spins to Strings to Membranes},'' {\em
  JHEP} {\bf 2021} (2020), no.~5 \href{https://arxiv.org/abs/2011.05728}{{\tt
  2011.05728}}.

\bibitem{Chester:2018aca}
S.~M. Chester, S.~S. Pufu, and X.~Yin, ``{The M-Theory S-Matrix From ABJM:
  Beyond 11D Supergravity},'' {\em JHEP} {\bf 2018} (2018), no.~8
  \href{https://arxiv.org/abs/1804.00949}{{\tt 1804.00949}}.

\bibitem{Alday:2022rly}
L.~F. Alday, S.~M. Chester, and H.~Raj, ``{M-theory on AdS$_{4}$\texttimes{}
  S$^{7}$ at 1-loop and beyond},'' {\em JHEP} {\bf 11} (2022) 091,
  \href{https://arxiv.org/abs/2207.11138}{{\tt 2207.11138}}.

\bibitem{Chang:2017cdx}
C.-M. Chang, M.~Fluder, Y.-H. Lin, and Y.~Wang, ``{Spheres, Charges,
  Instantons, and Bootstrap: A Five-Dimensional Odyssey},'' {\em JHEP} {\bf
  2018} (2017), no.~3 \href{https://arxiv.org/abs/1710.08418}{{\tt
  1710.08418}}.

\bibitem{Zhou:2017zaw}
X.~Zhou, ``{On Superconformal Four-Point Mellin Amplitudes in Dimension
  $d>2$},'' {\em JHEP} {\bf 2018} (2017), no.~8
  \href{https://arxiv.org/abs/1712.02800}{{\tt 1712.02800}}.

\bibitem{Chester:2018lbz}
S.~M. Chester, ``{AdS$_4$/CFT$_3$ for Unprotected Operators},'' {\em JHEP} {\bf
  2018} (2018), no.~7 \href{https://arxiv.org/abs/1803.01379}{{\tt
  1803.01379}}.

\bibitem{Alday:2018kkw}
L.~F. Alday, ``{On Genus-one String Amplitudes on $AdS_5 \times S^5$},''
  \href{https://arxiv.org/abs/1812.11783}{{\tt 1812.11783}}.

\bibitem{gurobi}
{Gurobi Optimization, LLC}, ``{Gurobi Optimizer Reference Manual},'' 2021.

\bibitem{Simmons-Duffin:2015qma}
D.~Simmons-Duffin, ``{A Semidefinite Program Solver for the Conformal
  Bootstrap},'' {\em JHEP} {\bf 06} (2015) 174,
  \href{https://arxiv.org/abs/1502.02033}{{\tt 1502.02033}}.

\end{thebibliography}\endgroup

\end{document}